\pdfoutput=1 
\documentclass{article}
\usepackage{arxiv}

\usepackage[utf8]{inputenc} 
\usepackage[T1]{fontenc}    
\usepackage[hidelinks]{hyperref}       
\usepackage{url}            
\usepackage{booktabs}       
\usepackage{amsfonts}       
\usepackage{nicefrac}       
\usepackage{microtype}      
\usepackage{lipsum}		

\usepackage{amssymb,amsmath,subfig,color,mathrsfs,lmodern}
\usepackage{graphicx}
\usepackage{cases}
\usepackage{natbib}
\usepackage{caption}
\usepackage{subfig}

\usepackage{url}

\usepackage{algorithm}
\usepackage{algpseudocode}
\algrenewcommand{\algorithmiccomment}[1]{\hskip1em\textit{$//$#1}}

\newcommand{\be}{\begin{equation}} \newcommand{\ee}{\end{equation}}
\renewcommand{\ee}{\end{equation}}
\newcommand{\bes}{\begin{equation*}}
\newcommand{\ees}{\end{equation*}}
\renewcommand{\vec}[1]{\boldsymbol #1} 

\renewcommand{\emph}{\textit}

\usepackage{soul,ulem,color,xspace,bm}
\renewcommand{\vec}[1]{\boldsymbol{#1}}

\newcommand{\dsp}{\mathbin{:}}

\newcommand{\ptf}{p^{\text{tf}}}
\newcommand{\ttf}{\text{tf}}

\renewcommand{\vec}[1]{\boldsymbol{#1}}
\newcommand{\ten}[1]{\boldsymbol{#1}}

\newcommand{\act}{\Gamma^\text{c}}

\newcommand{\tp}{^\intercal}

\title{Computational framework for monolithic coupling for thin fluid flow in contact interfaces}

\usepackage{authblk}

\author[a,b\thanks{Corresponding author: \tt{andrei.shvarts@glasgow.ac.uk}}]{Andrei G.~Shvarts}
\author[c]{Julien~Vignollet}
\author[a]{Vladislav A.~Yastrebov}

\affil[a]{\textit{Centre des Mat\'{e}riaux, CNRS UMR 7633, MINES ParisTech, PSL University, BP 87, 91003 Evry, France}}
\affil[b]{\textit{Glasgow Computational Engineering Centre, James Watt School of Engineering, University of Glasgow,\newline G12 8QQ Glasgow, United Kingdom}}
\affil[c]{\textit{Safran Tech, Safran Group, 1 rue Genevi\`{e}ve Aube, 78114 Magny-les-Hameaux, France}}

\date{}

\let\oldmaketitle\maketitle
\renewcommand{\maketitle}{\oldmaketitle\setcounter{footnote}{0}}

\renewcommand{\headeright}{}
\renewcommand{\undertitle}{}
	
\begin{document}
\maketitle
	
\begin{abstract}
We developed a computational framework for simulating thin fluid flow in narrow interfaces between contacting solids, which is relevant for a range of engineering, biological and geophysical applications. The treatment of this problem requires coupling between fluid and solid mechanics equations, further complicated by contact constraints and potentially complex geometrical features of contacting surfaces. We developed a monolithic finite-element framework for handling mechanical contact, thin incompressible viscous flow and fluid-induced tractions on the surface of the solid, suitable for both one- and two-way coupling approaches. Additionally, we consider the possibility of fluid entrapment in ``pools'' delimited by contact patches and its pressurization following a non-linear compressibility constitutive law. Image analysis algorithms were 
adapted to identify the local status of each interface element (i.e. distinguish between contact, fluid flow and trapped fluid zones) within the Newton-Raphson loop. First, an application of the proposed framework for a problem with a model geometry is given, and the robustness is demonstrated by the residual-wise and status-wise convergence. The full capability of the developed two-way coupling framework is demonstrated on a problem of a fluid flow in contact interface between a solid with representative rough surface and a rigid flat. 
The evolution of the contact pressure, fluid flow pattern and the morphology of trapped fluid zones under increasing external load until the complete sealing of the interface is displayed. 
Additionally, we demonstrated an almost mesh-independent result of a refined post-processing approach to the real contact-area computation.		
The developed framework permits not only to study the evolution of effective properties of contact interfaces, such as transmissivity and real contact area, but also to highlight the difference between one- and two-way coupling approaches, and, in particular, to quantify the effect of trapped fluid ``pools'' on the coupled problem.
\end{abstract}


\section{Introduction}

The problem of thin fluid flow in narrow interfaces between contacting or slightly separated deformable solids appears in different contexts: in engineering and biological applications, as well as in geophysical sciences. Rigorous handling of such problems requires solving a strongly nonlinear contact problem, which is further complicated by a multi-field coupling of essentially interrelated fluid and solid mechanics.  The free volume\footnote{By free volume, here, we mean the void separating the contacting surfaces.} between contacting surfaces depends on their initial geometry~\citep{paggi2015evolution}, which can be rather complex: they may have deterministic features, such as turned~\citep{rafols2016modelling}, patterned surfaces~\citep{sahlin2010mixed_b,prodanov2013tl}, and at a certain magnification may be considered as randomly rough~\citep{nayak1971tasme} down to atomistic scale~\citep{krim1995ijmpb,ponson2006ijf,whitehouse2010handbook,thomas1999b}.

Numerous applications of the problem of thin fluid flow in contact interfaces include sealing engineering~\citep{muller1998b}, lubrication in elasto-hydrodynamic and mixed regimes~\citep{stupkiewicz2009elastohydrodynamic,sahlin2010mixed_a} and functioning of human joints~\citep{caligaris2008effects}. Such an interaction between fluids and solids in contact is also relevant for hydraulic fracturing~\citep{bazant2014jam}, extraction of shale gas and oil from rocks~\citep{hubbert1972mechanics}, and at larger scales in landslides~\citep{viesca2012jgr}, slip in pressurized faults~\citep{garagash2012jgr} and bazal sliding of glaciers~\citep{fischer1997aog}.

\subsection{Fluid-structure interaction approaches}

The problem under discussion belongs to a vast domain of fluid-structure interaction (FSI) problems, which involve deformation and/or motion of solids interacting with internal and/or external fluid(s). These problems are of a very wide range, spanning from deformation of aeroplane wings and rotor blades subjected to the sub- or supersonic air flow~\citep{farhat2003application, bazilevs20113d} to modelling of the blood flow~\citep{bazilevs2006isogeometric, gerbeau2003quasi} and heart valves~\citep{dos2008partitioned,astorino2009fluid,kamensky2015immersogeometric,hiromi2020interface}, scaling up to suspended bridge instabilities under wind load~\citep{paidoussis2010fluid}, ship stability~\citep{wackers2011free} or huge iceberg's capsize in water~\citep{bonnet2020modelling}. All these problems correspond to different length and time scales, operating conditions and other requirements, therefore a unified FSI approach fit for all cases does not exist, and different techniques have been developed for particular problems. 

Many problems of the fluid-structure interaction, such as aeroelasticity and hemodynamics, correspond to the case of the high-Reynolds-number flow. Therefore, different mesh density, and often different time stepping, are required for the solid and fluid domains. Furthermore, the fluid domain evolves due to motion and deformation of solids. A number of methods have been used to overcome the associated computational complexity, such as arbitrary Lagrangian--Eulerian method~\citep{donea1982arbitrary,takashi1992arbitrary,wick2014flapping}, fictitious domain method~\citep{baaijens2001fictitious,dos2008partitioned,kadapa2016fictitious}, immersed boundary method~\citep{mittal2005immersed,kadapa2017stabilised,kadapa2018stabilised} and extended finite element method~\citep{mayer20103d,gerstenberger2008extended}, to name a few. 

On the contrary, fluid flow in contacting or slightly separated interfaces is usually of low Reynolds number and, moreover, the thickness of the fluid film is usually much smaller than other dimensions of the solid. In this case general Navier-Stokes equations could be readily simplified down to Reynolds equation for the viscous flow~\citep{Hamrock_2004}. This simplification permits to use compatible meshes for the fluid and the solid domains and, under assumption of constant pressure through the film thickness, to define the Reynolds equation on the so-called lubrication surface, so that specific methods discussed above are not required~\citep{stupkiewicz2009elastohydrodynamic,stupkiewicz2016finite}.

\subsection{One-way and two-way coupling approaches}

From the point of view of underlying physical processes, FSI strategies could be divided into one-way and two-way coupling approaches. In the context of thin fluid flow through contact interfaces, the former implies that the solution of the solid mechanics problem defines the distribution of the free volume in the interface which can be occupied by the fluid flow; however, the fluid pressure does not affect the deformation of the solids, i.e. the fluid problem is solved assuming rigid walls of solids. In the two-way coupling this approximation is dropped, and the effect of the fluid-induced traction acting on the surface of the deformable solid is taken into account.

In elastohydrodynamic lubrication regime, as well as for non-contact seals, two-way coupling is often used~\citep{stupkiewicz2009elastohydrodynamic,stupkiewicz2016finite,yang2009mortar}. 
However, for the important case of contact seals, or, more generally, if contact is present in the interface, the one-way coupling is rather utilized~\citep{dapp2012prl,dapp2016fluid,rafols2016modelling}. It is widely assumed in this context that the deformation of the solids happens mainly due to the contact interaction, and the fluid pressure effect on the solid is negligible, since the contact pressure at surface's asperities is considerably higher than the physically relevant fluid pressure. However, to the best of authors knowledge, a quantitative range of validity of one-way coupling for problems involving thin fluid flow in contact interfaces, depending on the surface geometry, material properties and the fluid pressure, has not been determined yet. The lack of such quantitative analysis is probably caused by insufficiency of existing numerical methods for comparison of one- and two-way coupling approaches for this type of problems, which is the main motivation of the current study.

It is important to note, that the aforementioned general FSI approaches to problems involving solid-to-solid contact enable two-way coupling, see e.g. \citep{mayer20103d,kamensky2015immersogeometric,kadapa2018stabilised}. However, these methods become inefficient when a very fine discretization is required  to accurately represent surface roughness in a numerical simulation. This computational complexity can be overcome by considering surface roughness in an averaged sense at the macroscopic level~\citep{patir1978average,perez2016stochastic,zaouter2018gas,waseem2017stochastic}, or as was done recently using a porous flow model~\citep{ager2019consistent}. Nonetheless, if one is interested in the effect of rough or deterministic surface features on the solution of the coupled problem, e.g. spatial distribution of contact stresses and fluid flow patterns, this approximation is not appropriate. Therefore, the main purpose of the current study is to develop a computational framework suitable for both one- and two-way coupling approaches and applicable for a discretization which reflects salient features of the surface geometry.

An important example of the effect of the surface roughness on the coupled problem, relevant only for the two-way coupling, is the phenomenon of fluid entrapment. Studying the evolution of the morphology of the contact area under increasing normal load, one may observe how non-simply connected contact patches appear, see~\citep{yastrebov2015ijss,lorenz2010time}. At the same time, the fluid present in the interface can be trapped inside ``pools'' (or ``valleys'') bounded by these patches and subsequently become pressurized, providing additional load-bearing capacity. 
The behaviour of trapped fluid accounts for a significant reduction of friction in tyre--road contact~\citep{scaraggi2012time}, cold metal forming~\citep{azushima1995direct} and functioning of human joints~\citep{soltz2003hydrostatic,chan2011role}.  However, the effect of the fluid entrapment on transmissivity of contact interfaces, which is an import characteristic for sealing engineering, has not been thoroughly investigated yet.

Recently, we studied the behaviour of pressurized fluid trapped in a contact interface without considering the fluid flow~\citep{shvarts2018trapped}. Using the finite element framework, we successfully introduced a novel trapped fluid superelement based on all out-of-contact segments (faces in 3D), surrounded by a contact patch. Therefore, an additional objective and a major novelty of the current study is the incorporation of such a formulation of the trapped fluid element into  a monolithic computational framework for coupling contact and fluid equations, which will permit to quantify the effect of trapped fluid zones on the solution of the problem. 

\subsection{Partitioned and monolithic implementation approaches}

From the implementation point of view two distinct approaches for any FSI problem exist: partitioned and monolithic, see e.g.~\citep{felippa2001partitioned}. The former is based on two different solvers for the fluid and solid sub-problems, and in order to take into account the coupling, one- or two-way data exchange between them must be established, while a certain iterative process may be required to obtain the convergence. At the same time, the utilization of the partitioned approach benefits from modularity, since different solvers tailored for the sub-problems could be used~\citep{kuttler2008fixed,matthies2003partitioned}. On the contrary, under the \textit{monolithic} approach all equations which govern sub-problems and the interaction between them are rendered into a single system~\citep{hubner2004monolithic,michler2004monolithic,heil2004efficient,verdugo2016unified}. Upon its solution DOF values corresponding to both sub-problems are obtained simultaneously, therefore, the data exchange in this case is not needed. 

Elasto-hydrodynamic lubrication problems are often solved under the monolithic approach~\citep{stupkiewicz2016finite,stupkiewicz2009elastohydrodynamic,yang2009mortar}, whilst for the contact sealing problems the partitioned approach is generally preferred~\citep{dapp2012prl,rafols2016modelling}. Moreover, as was already mentioned above, the latter problem is often solved under the one-way coupling approach, using the assumption of the infinitesimal slopes of the surface profile and the small deformation formulation. Boundary element method~\citep{rafols2016modelling} and Green's function molecular dynamics~\citep{dapp2012prl} are frequently used for the mechanical contact problem and the Reynolds equation is often solved by the finite-differences method.

\subsection{Objectives and outline}
In this paper we develop a computational framework aimed first at solving the two-way coupling of the mechanical contact and fluid flow sub-problems, which could require frequent and considerable data exchange in case of the partitioned approach. The second objective is to take into account the effect of trapped fluid pools appearing in the interface. If the number of these zones becomes large, then a resolution under the partitioned approach becomes even more complicated, since the history tracking of trapped zones is needed. The \textit{monolithic} approach appears beneficial for our purposes and, therefore, it was applied throughout this study. 
We use the finite-element method for implementing the monolithic approach. This choice also makes possible the application of the developed tool to different surface geometries (including surfaces with finite slopes), under large deformation formulation and with different material models of the solid (e.g. elasto-plastic, viscoelastic, etc). 

Note that the methodology presented here was already used to solve a two-way coupling problem of a pressure-driven fluid flow in contact interface between an elastic solid with an extruded wavy surface and a rigid flat~\citep{shvarts2018fluid}. A good agreement was observed between the numerical solution and an approximate analytical formula (derived in the same paper), in the interval of loads within which the latter is applicable.
However, since only regular wavy profiles were studied, the problem did not include trapped fluid zones, therefore a complete testing of the proposed method was not possible. Moreover, the methodology was only used, but was not presented in detail in the above cited work.

The present paper is organized as follows. In Section 2 we formulate the coupled problem by outlining equations governing sub-parts for each physics, while Section 3 introduces the variational statement for this problem under both one- and two-way coupling approaches. In Section 4 we propose the monolithic framework and provide ready-to-implement finite-element formulations of the tangent matrix and the residual vector of the coupled problem. Moreover, we discuss the algorithm used to determine the local status in the interface and keep track of formation and evolution of multiple trapped fluid zones. Section 5 is devoted to examples showing capabilities of the proposed framework and some relevant discussions. Finally, Section 6 provides a short conclusion, while the Appendix~\ref{sec:appendix} outlines detailed expressions for components of residual vectors and tangent matrices of discussed elements.

\section{Coupled problem statement}
\label{sec:coupled_problem_statement}
We consider a problem of a thin fluid flow in contact interface formed between a solid body with the surface geometry given for concreteness by function $z(x, y)$ and a rigid flat described by plane $z=0$. This set-up was chosen to simplify the formulation of the contact problem and to concentrate the discussion on handling of the fluid/solid coupling. Furthermore, for the same reason, we will consider only small deformation/small rotation formulation; justifying arguments for these assumptions regarding the problem of thin fluid flow in contact interfaces will be given below.  

Note that the unilateral contact set-up is equivalent to the case of contact between two deformable bodies under linear isotropic elasticity and infinitesimal strain assumptions~\citep{barber2003bounds}. Furthermore, the problem statement with one of the contacting solids considered as rigid is relevant when the Young's modulus of this solid is considerably higher than this of the other, which is the case, for example, in metal-to-rubber sealing applications~\citep{persson2008theory} and in tyre--road contact~\citep{scaraggi2012time}. 
At the same time, the numerical framework developed in this paper for the case of unilateral contact can be extended afterwards to include the problem of contact between two deformable solids with arbitrary surface geometries.

Let us denote by $\Omega$ a deformable solid and by $\Gamma \subset \partial \Omega$ a part of its surface which represents the \textit{potential} contact zone, i.e. defines the maximal possible extent of the contact interface. The resolution of the coupled problem for any given boundary conditions requires partitioning of the surface $\Gamma$ into subsets, according to the local status\footnote{By the local status in this context we mean that each point of the solid's surface may belong to either the active contact zone, or fluid-structure interface, or one of the trapped fluid zones.} of each point of the interface, see Fig.~\ref{fig:problem}:
\begin{equation}
\label{eq:prob:domain}
\Gamma = \Gamma^\text{c} \cup \Gamma^{\text{fsi}} \cup \bigcup\limits_{i=1}^{n_\ttf} \Gamma^{\text{tf}}_i.
\end{equation}
The notation of the subsets is the following: $\Gamma^\text{c}$ is the active contact zone (which may consist of several separated contact clusters), where normal contact tractions are non-zero, 
$\Gamma^{\text{fsi}}$ is a part of the solid's surface which interacts with the flowing fluid and where the surface tractions are equal to the corresponding tractions in the fluid (so-called \textit{fluid-structure interface}), and 
$\Gamma^{\text{tf}}_i, \: i = \overline{1, n_{\text{tf}}}$ are {\it trapped fluid} zones, i.e. parts of the surface $\Gamma$ which are out of contact, but completely delimited by non-simply connected contact patches, {which encircle pools containing the fluid. Furthermore, we term by $\Gamma^\text{f}$ the projection of $\Gamma^{\text{fsi}}$ on the plane $z=0$, which serves as the lubrication surface where the Reynolds equation for the fluid flow will be defined. 

By definition, $\Gamma^\text{c} \cap \Gamma^\text{fsi} = \emptyset$ and $\Gamma^\text{c} \cap\Gamma^{\ttf}_i = \emptyset \; \forall i = \overline{1, n_{\text{tf}}}$. Note also, that $\Gamma^\text{fsi} \cap \Gamma^{\ttf}_i = \emptyset \; \forall i = \overline{1, n_{\text{tf}}}$, i.e. even though all trapped pockets contain the same fluid as the one present in the fluid-flow domain, the behaviour of the trapped fluid is to be considered separately.
We assume here that in the initial configuration $\Gamma^\text{c} = \emptyset$ and $n_{\ttf} = 0$, so that the fluid flow zone occupies the whole interface. This assumption makes impossible appearance of non-contact zones which do not belong to $\Gamma^{\text{fsi}}$ or to one of the trapped fluid zones $\Gamma^{\ttf}_i$. However, consideration of such a case (e.g., air bubbles entrapment in the case of non-saturated fluid interface) could be included into the framework. 

\begin{figure}[t]	
	\centering
	\includegraphics[width=0.6\textwidth]{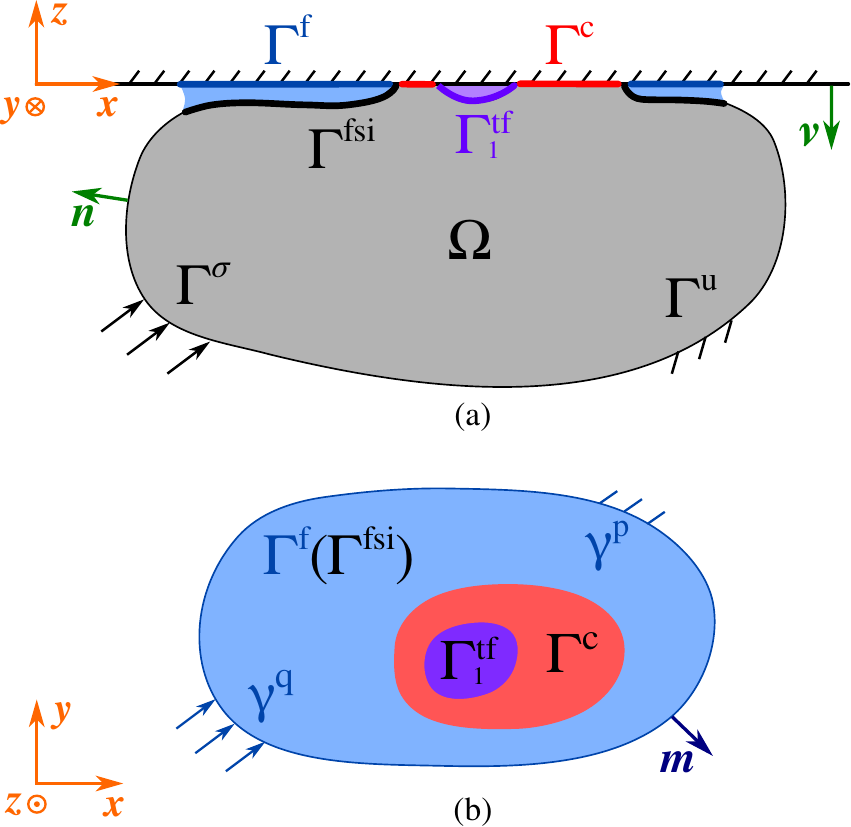}
	\caption{Sketch of the problem under study: (a) contact between a solid $\Omega$ and a rigid flat with fluid present in the interface; (b) view of the interface $\Gamma$. Notations: $\Gamma^\text{c}$ is the active contact zone, $\Gamma^{\text{fsi}}$ is the fluid-structure interface, its projection on the rigid flat is denoted as $\Gamma^\text{f}$, which is the lubrication surface where the Reynolds equation is solved, and $\Gamma^{\text{tf}}_1$ is the trapped fluid zone.}
	\label{fig:problem}
\end{figure}

\subsection{Solid mechanics problem with unilateral contact}	

The deformation of the solid (in absence of the fluid) is governed by the balance of momentum equation complemented by the contact and boundary conditions, which in absence of inertial and volumetric forces reads as:
\begin{subnumcases}{\label{eq:contact}}
\label{eq:balance}\nabla\cdot\ten{\sigma}^\text{s}(\vec{u}) = 0 & in $\Omega$ \\
\label{eq:hsm}g(\vec{u}) \geq 0, \; \sigma_n(\vec{u}) \leq 0, \; g(\vec{u}) \: \sigma_n(
\vec{u}) = 0 & on $\Gamma$ \\
\label{eq:bc_dirichlet_s} \vec{u} = \vec{u}_0 & on $\Gamma^u$ \\
\label{eq:bc_neumann_s} \ten{\sigma}^\text{s} \cdot \vec{n} = \vec{t}_0 & on $\Gamma^{\sigma}$ 		
\end{subnumcases}			
where $\vec{u}$ is the displacement field, $\ten{\sigma}^\text{s}$ is the Cauchy stress tensor;~\eqref{eq:hsm} are the Hertz--Signorini--Moreau conditions for non-adhesive frictionless unilateral contact~\citep{wriggers2006b,yastrebov2013b}, $\sigma_n=\vec{n}\cdot\ten{\sigma}^\text{s}\cdot\vec{n}$ is the normal traction on the solid's surface (to which $\vec{n}$ is the outer normal). For each point $\vec{x}\in\Gamma$, $\vec{x} = [x,y,z]^\intercal$, we denote by $g(\vec{u})$ the normal gap, i.e. the signed distance between the surface of the solid and the rigid flat: $g>0$ in case of separation, $g=0$ in contact and $g<0$ refers to non-admissible penetration:
\begin{equation}
\label{eq:prob:gap}
g(\vec{u}) = g_0 + \vec{u}\cdot\vec{\nu},
\end{equation} 
where $g_0$ is the initial gap of this point and $\vec{\nu}$ is the outward normal to the rigid flat. 
Finally, \eqref{eq:bc_dirichlet_s} are the Dirichlet boundary
conditions with a prescribed displacement $\vec{u}_0$ and \eqref{eq:bc_neumann_s} are the Neumann boundary conditions with a prescribed surface traction $\vec{t}_0$, defined on $\Gamma^u\subset\partial\Omega$ and $\Gamma^\sigma\subset\partial\Omega$, respectively. 

Note that our coupled framework concerns processes occurring in the contact interface only 
and permits arbitrary constitutive laws for the underlying solid, see examples in~\citep{shvarts2019phd}. 
Here, in order to simplify the discussion and concentrate it on the aspects of the coupling, we will  consider isotropic linearly elastic solid, i.e. the Cauchy stress tensor is related to the infinitesimal strain tensor $\ten\varepsilon = (\nabla\vec{u} + (\nabla\vec{u})^\intercal)/2$ by Hooke's law: 
$\ten\sigma^\text{s} = \lambda\; \mathrm{trace}(\ten\varepsilon)\ten I + 2\mu\ten\varepsilon$ 
with the Lam\'{e} constants (elastic moduli) $\lambda$ and $\mu$, $\ten I$ is the identity tensor. 

\subsection{Thin fluid flow}

Working under assumptions of thin laminar flow and neglecting inertial and volumetric forces, Navier-Stokes equations, which govern the mechanics of a viscous fluid in a general case, can be reduced to the Reynolds equation, see e.g.~\citep{Hamrock_2004}, without loss of accuracy for the class of applications considered in this paper. Accordingly, we state the boundary value problem for the thin fluid flow as follows:
\begin{subnumcases}{\label{eq:fluid_flow}}
\label{eq:reynolds}\nabla\cdot\left[g(\vec{u})^3 \nabla p\right] = 0 & in $\Gamma^\text{f}$ \\
\label{eq:bc_dirichlet_f} p = p_0 & on $\gamma^p$ \\
\label{eq:bc_neumann_f} \vec{q}\cdot\vec{m} = q_0 & on $\gamma^q$ 
\end{subnumcases}
where~\eqref{eq:reynolds} is the Reynolds equation for isoviscous incompressible Newtonian fluid, note that the tangential relative motion of the solid walls is not considered here whereas the normal approaching is assumed to be quasi-static, $p(x,y)$ is the fluid pressure field defined on the lubrication surface $\Gamma^\text{f}$, which is a projection of the fluid-structure interface $\Gamma^{\text{fsi}}$ on the plane $z=0$, and the $\nabla$ operator in~\eqref{eq:reynolds} is defined as $\nabla(\cdot) := \partial(\cdot)/\partial x\,\vec{e}_x + \partial(\cdot)/ \partial y\, \vec{e}_y$, where $\{\vec e_x,\vec e_y\}$ are orthonormal basis vectors lying in the plane $z=0$. Finally, \eqref{eq:bc_dirichlet_f} are the Dirichlet boundary conditions with a prescribed fluid pressure $p_0$ and \eqref{eq:bc_neumann_f} are the Neumann boundary conditions with a prescribed fluid flux $q_0$, defined on $\gamma^p\subset\partial\Gamma^\text{f}$ and $\gamma^q\subset\partial\Gamma^\text{f}$, respectively ($\vec{m}$ is the outward normal to $\Gamma^\text{f}$), while the fluid flux is given by: 
\begin{equation}
\label{eq:prob:flux}
\vec q =  -\frac{g(\vec{u})^3}{12\mu} \nabla p,
\end{equation}
where $\mu$ is the dynamic viscosity. Note that for each point $(x,y)\in\Gamma^\text{f}$ the thickness of the film is computed as the normal gap $g(\vec{u})$ of the corresponding point $(x,y,z)\in\Gamma^\text{fsi}$.

It is important to note, that in addition to conditions outlined above, an assumption of sufficiently small slopes of the surface geometry is required for the validity of the Reynolds equation. Omitting this assumption results in inconsistencies between predictions of the Reynolds equation and those of the full Navier-Stokes equations, for example, in the problem of the fluid flow through fractures~\citep{brown1995applicability}. At the same time, consideration of small slopes of the surface geometry justifies the assumption of small deformations and small rotations for the contact problem discussed above, see also~\citep{barber2003bounds}.

\subsection{Fluid-structure interface}
We do not need to consider the no-slip condition (i.e. zero flow velocity at the fluid-structure boundary), as, for example, in~\citep{farhat1998load}, since it is already taken into account in the considered form of the Reynolds equation~\eqref{eq:reynolds}. At the same time, the equilibrium of the solid and fluid tractions on the fluid-structure interface $\Gamma^\text{fsi}$ needs to fulfil the following equation:
\begin{equation}
\label{eq:fsi_f} \ten{\sigma}^\text{s} \cdot \vec{n} = -p \: \vec{n} -\frac{g(\vec{u})}{2} \nabla{p} \quad \text{on} \; \Gamma^\text{fsi},
\end{equation}
where the first right-hand side term is the normal traction due to the hydrostatic pressure, while the second one is the tangential traction due to shear stresses in the fluid that act on the solid's surface (here, it results from the Poiseuille flow), see~\citep{Hamrock_2004} for details.

Note that the $\nabla$ operator here is defined on the lubrication surface $\Gamma^\text{f}$, and the second right-hand side term in~\eqref{eq:fsi_f} is not exactly perpendicular to the outward normal $\vec{n}$ to the surface $\Gamma^\text{fsi}$.  Nevertheless, this slight inconsistency is justified by the requirement of small slopes of the surface geometry mentioned above, and is often accepted in elasto-hydrodynamic lubrication problems~\citep{stupkiewicz2009finite,stupkiewicz2016finite}.
 
Moreover, in the derivation of the Reynolds equation~\eqref{eq:reynolds} the thickness of the fluid film is assumed to be much smaller than other length scales, and therefore, the term corresponding to the tangential traction in~\eqref{eq:fsi_f} is often neglected in lubrication problems. 
However, in application to sealing problems, studies of the elasto-hydrodynamic lubrication regime show a noticeable effect of the shear tractions on the seal's leakage, see~\citep{stupkiewicz2009elastohydrodynamic}.
Therefore, for the sake of completeness, we will consider both normal and shear components of the fluid-induced traction in the developed framework. 

\subsection{Trapped fluid zones}

The hydrostatic pressure $\ptf_i$, developed in the $i$-th trapped fluid zone, is applied to the surface of the solid body as the normal traction:
\begin{equation}
\label{eq:coup:trap_press} \sigma_n = -\ptf_i \quad \text{on} \; \Gamma^{\ttf}_i, \quad i = \overline{1, n_{\text{tf}}}.
\end{equation} 
However, the pressure $\ptf_i$ is \textit{a priori} unknown, and the behaviour of the trapped fluid is not governed by the Reynolds equation. Furthermore, the pressure developed in such trapped fluid pocket can be considerably higher than the pressure in the fluid flow, and therefore a model of \textit{incompressible} (or even linearly compressible) fluid becomes inaccurate, see discussions in~\citep{shvarts2018trapped}. Therefore, we will consider a model of \textit{compressible} fluid with a pressure-dependent bulk modulus $K = K_0 + K_1\:\ptf_i$, where $K_0$ [Pa] and $K_1$ (dimensionless) are model parameters, suitable for fluids typically used in lubrication and sealing applications, see also~\citep{kuznetsov1985effect} for details. According to this model, the pressure of the trapped fluid is a non-linear function of the relative change of its volume:
\begin{equation}
\label{eq:coup:nonlin_comp} \ptf_i =  \left(\ptf_{i0}+\frac{K_0}{K_1}\right) \left(\frac{V^\ttf_i}{V^\ttf_{i0}}\right)^{-K_1} - \frac{K_0}{K_1},
\end{equation}
where $V^\text{tf}_i$ is the current volume of fluid in the $i$-th zone, $V^\text{tf}_{i0}$ is the volume of the fluid in this zone at the moment when it was formed, and $p^{\ttf}_{i0}$ is the corresponding initial pressure of this trapped fluid, see~\citep{shvarts2018trapped} for more details.

Therefore, since the behaviour of each trapped fluid zone depends on its own set of initial parameters ($V^\text{tf}_{i0}$ and $p^{\ttf}_{i0}$), each zone is to be considered separately from others. 
As was mentioned above, we assume that in the initial configuration the fluid occupies the whole free volume between the contacting surfaces. 
Accordingly, the volume of the fluid in the $i$-th pool is equal to the volume of the gap between the surface $\Gamma^{\ttf}_i$ and the rigid flat, i.e. the fluid fills the entire volume of the closed pocket:
\begin{equation}
\label{eq::coup:gap_vol}
V^\text{tf}_i = \int\limits_{\Gamma^{\ttf}_i}g(\vec{u})(-\vec{n}\cdot\vec{\nu}) \: d\Gamma,
\end{equation}
where $\vec{n}$ is the outward normal to the surface $\Gamma^{\ttf}_i$, and $\vec{\nu}$ is the normal to the rigid flat, see Fig.~\ref{fig:problem}.

It is important to note, that the presented problem set-up corresponds to the two-way coupling approach, and therefore not only the displacement field $\vec{u}$ depends on the fluid pressure $p$ and \textit{vice versa}, but also the extent of fluid-flow domain $\Gamma^\text{f}$ and trapped fluid zones $\Gamma^{\text{tf}}_i$ depends on the morphology of the active contact zone, i.e. on the resolution of contact constraints.  Additional effort may be necessary for handling edge effects, e.g. enforcing continuity of surface tractions across boundaries between contact and fluid-flow zones, and also between contact and trapped fluid zones. Below we will discuss in detail our recipes of partitioning the interface and handling these issues. At the same time, the one-way coupling approach, which neglects the action of the fluid pressure on surface of the solid, can also be considered in the present problem statement if equations~\eqref{eq:fsi_f}-\eqref{eq::coup:gap_vol} are omitted.

\section{Variational formulation of the coupled problem}
\label{sec:var}
Before presenting the numerical framework, we discuss the variational formulation of the coupled problem stated in the previous section. We start by outlining contribution of each sub-problem to the balance of virtual work and then provide the variational formulation of the coupled problem for both one- and two-way coupling approaches.

\subsection{Weak formulation of the solid mechanics problem with contact constraints}
\label{sec:weak:cont}

The variational statement of the solid mechanics problem with contact constraints~\eqref{eq:contact}, see e.g.~\citep{kikuchi1988contact}, consists in finding a function $\vec{u}\in\mathbf{K}$:
\begin{equation}
\label{eq:cont:subsp}
\mathbf{K} = \left\{\left.\vec{u}\in\mathbf{V}\;\right| \; g(\vec{u}) \geq 0 \;\text{on}\; \Gamma\right\}, \quad
\mathbf{V} = \left\{\left.\vec{u}\in\mathbf{H}^1\left(\Omega\right)\;\right|\,\vec{u} = \vec{u}_0 \;\text{on}\; \Gamma^u\right\},
\end{equation}
such that:
\begin{equation}
\label{eq:cont:virt_work_ineq}
\int\limits_{\Omega} \ten{\sigma}^\text{s}(\vec{u})\dsp\nabla\delta\vec{u}\,d\Omega - \int\limits_{\Gamma^{\sigma}}\vec{t}_0\cdot\delta\vec{u}\,d\Gamma \ge 0 \quad \forall\,\delta\vec{u}\in\mathbf{V_0},
\end{equation}
where $\mathbf{V_0}$ is a subspace for virtual displacements: 
\begin{equation}
\label{eq:cont:lin_sp}
\mathbf{V_0} = \left\{\left.\vec{v}\in\mathbf{H}^1\left(\Omega\right)\;\right|\:\vec{v} = \vec{0} \;\text{on}\; \Gamma^u\right\},
\end{equation}
and $\mathbf{H}^1\left(\Omega\right)$ denotes the Sobolev space for vector-valued functions with square-integrable derivatives.

The variational statement of the problem~\eqref{eq:contact} takes the form of the inequality~\eqref{eq:cont:virt_work_ineq} instead of a standard (equality) variational principle due to the presence of contact constraints defining a subset of admissible displacements $\mathbf{K} \subset \mathbf{V}$, see~\eqref{eq:cont:subsp}. In order to transform~\eqref{eq:cont:virt_work_ineq} to a standard variational principle, we use the augmented Lagrangian method, which combines benefits of the classic Lagrange multipliers method (exact satisfaction of the constraints) and the penalty method (rendering the functional smooth across the transition of contact states), see also~\citep{alart1991mixed}. Thus, we introduce the augmented Lagrangian for the problem~\eqref{eq:contact}:
\begin{equation}
	\label{eq:aug_lag}
	\mathcal{L}^\text{a} (\vec{u},\lambda) = \frac{1}{2}\int\limits_{\Omega}\ten{\sigma}^\text{s}(\vec{u})\dsp\nabla\vec{u}\,d\Omega - \int\limits_{\Gamma^{\sigma}}\vec{t}_0\cdot\vec{u}\,d\Gamma+\int\limits_{\act} \left[\lambda\,g(\vec{u}) + \frac{\epsilon}{2}\,g(\vec{u})^2\right]d\Gamma - \int\limits_{\Gamma\setminus\act} \frac{1}{2\epsilon}\,\lambda^2 \, d\Gamma,
\end{equation}
where the first two terms constitute the potential energy $\Pi^\text{s}(\vec{u})$ of the solid in absence of contact, and the two last terms correspond to the contact interaction energy $W^\text{c}(\vec{u},\lambda)$, computed for both active $\act$ and non-active $\Gamma\setminus\act$ zones of contact, see~\citep{yastrebov2013b} for more details. Here $\epsilon$ is the augmentation parameter, and $\lambda\in\mathcal{M}$ is the Lagrange multiplier function, equivalent to the normal traction $\sigma_n$ on the surface $\Gamma$: $\mathcal{M} \equiv H^{-\frac{1}{2}}(\Gamma)$, i.e. the space for Lagrange multipliers is dual to $H^{\frac{1}{2}}(\Gamma)$, which is the trace of the space $H^1(\Omega)$ on the boundary $\Gamma$, see~\citep{kikuchi1988contact}. 

The solution of the contact problem is found as the stationary saddle point of the Lagrangian~\eqref{eq:aug_lag}, at which its variation vanishes:
\begin{equation}
\delta\mathcal{L}^a (\vec{u},\lambda) = D\Pi^\text{s}(\vec{u}) \cdot \delta\vec{u} + D W^\text{c}(\vec{u},\lambda) \cdot \delta\vec{u} + D W^\text{c}(\vec{u},\lambda) \, \delta\lambda = 0, \quad \forall\,\delta\vec{u}\in\mathbf{V_0}, \; \forall\,\delta\lambda\in\mathcal{M},
\end{equation}
where we used the notation of directional (Gateaux) derivative, defined for the potential energy $\Pi^\text{s}(\vec{u})$ as
\begin{equation}
	D\Pi^\text{s}(\vec{u}) \cdot \delta\vec{u} = \lim_{\alpha\rightarrow 0}\frac{\Pi^\text{s}(\vec{u}+\alpha\,\delta\vec{u})-\Pi^\text{s}(\vec{u})}{\alpha},
\end{equation}
and defined similarly for the contact energy term $W^\text{c}(\vec{u},\lambda)$, see~\citep{wriggers2006b} for more details. Therefore, the variational inequality~\eqref{eq:cont:virt_work_ineq} is transformed into a problem of finding functions $\vec{u}\in\mathbf{V}$ and $\lambda\in\mathcal{M}$ such that they satisfy the variational equality: 
\begin{subequations}
 	\label{eq:cont:var}
 	\begin{alignat}{2}
	\label{eq:cont:var:disp}
	\int\limits_{\Omega} \ten{\sigma}^\text{s}(\vec{u})\dsp\nabla\delta\vec{u}\,d\Omega - \int\limits_{\Gamma^{\sigma}}\vec{t}_0\cdot\delta\vec{u}\,d\Gamma + \int\limits_{\act} \left[\lambda + \epsilon\,g(\vec{u})\right] \frac{\partial g(\vec{u})}{\partial\vec{u}} \cdot\delta\vec{u}\, d\Gamma &= 0, \quad \forall\,\delta\vec{u}\in\mathbf{V_0},\\
	\label{eq:cont:var:lag}
	\int\limits_{\act} g(\vec{u}) \,\delta\lambda\, d\Gamma - \int\limits_{\Gamma\setminus\act} \frac{1}{\epsilon}\,\lambda\,\delta\lambda\,d\Gamma &= 0, \quad \forall\,\delta\lambda\in\mathcal{M}.
	\end{alignat}
\end{subequations}

\subsection{Weak statement of the fluid flow sub-problem}
\label{ssec:weak_fluid_flow}
In order to obtain the weak form of the fluid-flow problem~\eqref{eq:fluid_flow}, we follow the standard approach to elliptic equations, see e.g.~\citep{zienkiewicz1977finite}, which results in the following statement: find a scalar field $p\in\mathcal{P}$:
\begin{equation}
\label{eq:var:fluid_space}
\mathcal{P} = \left\{\left.p\in H^1(\Gamma^\text{f})\;\right|\:p = p_0\;\text{on}\; \gamma^p\right\},
\end{equation}
such that
\begin{equation}
\label{eq:var:rey_int}
\int\limits_{\Gamma^\text{f}} \frac{g(\vec{u})^3}{12\eta}\,\nabla p\cdot\nabla\delta p \, d\Gamma + \int\limits_{\gamma^q} q_0 \,\delta p \, d\gamma = 0, \quad \forall\, \delta p \in \mathcal{P}_0,
\end{equation}
where $ \mathcal{P}_0$ is a subspace for scalar virtual functions:
\begin{equation}
\label{eq:var:fluid_space_var}
\mathcal{P}_0 = \left\{\left.\delta p\in H^1(\Gamma^\text{f})\;\right|\:\delta p = 0\;\text{on}\; \gamma^p\right\}.
\end{equation}

\subsection{Weak form of the fluid-structure interface balance}
\label{ssec:weak_fsi}
We may compute the work of the fluid-induced tractions on the surface $\Gamma^\text{fsi}$~\eqref{eq:fsi_f}, corresponding to a virtual displacement $\delta\vec{u}$, as: 
\begin{equation}
\label{eq:var:work_fsi}
\delta W^{\text{fsi}} = \int\limits_{\Gamma^{\text{fsi}}} \left(-p \vec{n} - \frac{g(\vec{u})}{2}\nabla p\right)\cdot\delta \vec{u} \: d \Gamma,
\end{equation}
and include it to Eq.~\eqref{eq:cont:var:disp} of balance of the virtual work with the negative sign, since the virtual work of surface tractions has a sign opposite to the one of the work of internal forces. It is also important to bear in mind that the traction in~\eqref{eq:var:work_fsi} is not known \textit{a priori} and depends on the fluid pressure $p$, its gradient $\nabla p$ and the displacement field $\vec{u}$ which determines the gap field. 

\subsection{Virtual work of trapped fluid zones}

To capture the effect of the trapped fluid pressure on the solid, we recall the thermodynamic definition of the elementary work done on a system, corresponding to an infinitesimal change of its volume. Following this concept, we compute the virtual work of an $i$-th trapped pool on the surface of the solid as:
\begin{equation}
\label{eq:weak:trap_1}
\delta W^\text{tf}_i = -p^\text{tf}_i\,\delta V^\text{tf}_i,
\end{equation}
where the minus sign is used since an increase of the volume of a trapped pool leads to a decrease of its pressure, and consequently, to a release of the energy of the trapped fluid. Since the volume of the fluid inside a trap $V_i$ is a functional of the displacement field $\vec{u}$ as defined by the integral~\eqref{eq::coup:gap_vol}, $\delta V_i$ can be treated as its first variation and computed using the directional (Gateaux) derivative: 
\begin{equation}
\label{eq:dir_der_vol}
\delta V^\text{tf}_i = D\,V^\text{tf}_i(\vec{u})\cdot\delta\vec{u} = \lim_{\alpha\rightarrow 0}\frac{V^\text{tf}_i(\vec{u}+\alpha\delta\vec{u}) - V^\text{tf}_i(\vec{u})}{\alpha}.
\end{equation} 
Therefore, substituting the non-linear formula for the trapped fluid pressure~\eqref{eq:coup:nonlin_comp}, the virtual work of a trapped fluid pocket corresponding to a virtual displacement $\delta\vec{u}$ can be expressed as:
\begin{equation}
\label{eq:weak:trap}
\delta W^\text{tf}_i = -\left[\left(\ptf_{i0}+\frac{K_0}{K_1}\right) \left(\frac{V^\ttf_i(\vec{u})}{V^\ttf_{i0}}\right)^{-K_1} - \frac{K_0}{K_1}\right] D\,V^\text{tf}_i(\vec{u})\cdot\delta{\vec{u}},
\end{equation} 
which can now be included into the equation for the virtual work~\eqref{eq:cont:var:disp}, taking into account the contribution of each trapped fluid zone separately.

\subsection{Variational formulation of the coupled problem}

Combining contributions of the sub-problems outlined above, we provide the variational statement of the coupled problem in the spirit of the monolithic approach~\citep{yang2009mortar,stupkiewicz2018finite}:
find fields $\vec{u} \in \mathbf{V}$, $\lambda \in \mathcal{M}$ and $p \in \mathcal{P}$ such that:
\begin{subequations}
	\label{eq:weak:coup_prob}
	\begin{alignat}{2}
	\label{eq:weak:solid} G^\text{s}(\vec{u}, \lambda, p; \delta\vec{u}) = 0,\quad  & \forall\:\delta\vec{u}\in\mathbf{V_0},\\
	\label{eq:weak:cont} G^\text{c}(\vec{u}, \lambda; \delta\lambda) = 0,\quad  & \forall\:\delta\lambda \in\mathcal{M},\\
	\label{eq:weak:fluid} G^\text{f}(\vec{u}, p; \delta p) = 0,\quad & \forall\:\delta p \in\mathcal{P}_0, 
	\end{alignat}
\end{subequations}
where 
\begin{subequations}\label{eqn:main}
	\begin{align}
	\label{eq:coup:solid:cont}G^\text{s}(\vec{u}, \lambda, p; \delta\vec{u}) &=\displaystyle\int\limits_{\Omega} \ten{\sigma}^\text{s}(\vec{u})\dsp\nabla\delta\vec{u}\,d\Omega  -\displaystyle\int\limits_{\Gamma^{\sigma}}\vec{t}_0\cdot\delta\vec{u}\,d\Gamma+
	\int\limits_{\act} \left[\lambda + \epsilon\,g(\vec{u})\right] \frac{\partial g(\vec{u})}{\partial\vec{u}} \cdot\delta\vec{u}\, d\Gamma\\
	\label{eq:coup:solid:fsi}\displaystyle&+\int\limits_{\Gamma^{\text{fsi}}} \left(p\,\vec{n} + \frac{g(\vec{u})}{2}\nabla p\right)\cdot\delta \vec{u} \, d \Gamma\\
	\label{eq:coup:solid:trap}\displaystyle&-\sum\limits_{i=1}^{n_\text{tf}}\left[\left(\ptf_{i0}+\frac{K_0}{K_1}\right)\left(\frac{V^\ttf_i(\vec{u})}{V^\ttf_{i0}}\right)^{-K_1} - \frac{K_0}{K_1}\right]D V^\text{tf}_i(\vec{u})\cdot\delta{\vec{u}},\\
	G^\text{c}(\vec{u}, \lambda; \delta\lambda) &= \int\limits_{\act} g(\vec{u}) \,\delta\lambda\, d\Gamma - \int\limits_{\Gamma\setminus\act} \frac{1}{\epsilon}\,\lambda\,\delta\lambda\,d\Gamma,\\
	\label{eq:coup:fluid}\displaystyle G^\text{f}(\vec{u}, p; \delta p) &= \int\limits_{\Gamma^\text{f}} \frac{g^3(\vec{u})}{12\eta}\,\nabla p\cdot\nabla\delta p \, d\Gamma + \int\limits_{\gamma^q} q_0 \,\delta p \, d\gamma.
	\end{align}
\end{subequations}

The weak problem statement~\eqref{eq:weak:coup_prob}-\eqref{eqn:main} is valid for the two-way coupling approach, when both sub-problems have an impact on each other. The one-way coupling approximation for the problem under study can also be considered upon following modifications: (\textit{i}) omit the fluid-induced tractions on the surface of the solid~\eqref{eq:coup:solid:fsi}, (\textit{ii}) neglect the effect of trapped fluid zones~\eqref{eq:coup:solid:trap}, and (\textit{iii}) assume rigid solid walls while solving fluid-flow equation~\eqref{eq:weak:fluid}. Therefore, in case of one-way coupling, instead of~\eqref{eq:weak:coup_prob}, we shall have the following equations:
\begin{subequations}
	\label{eq:weak:coup_one}
	\begin{alignat}{2}
	\label{eq:weak:solid_one} G^\text{s}(\vec{u}, \lambda; \delta\vec{u}) = 0,\quad  & \forall\:\delta\vec{u}\in\mathbf{V}_0,\\
	\label{eq:weak:cont_one} G^\text{c}(\vec{u}, \lambda; \delta\lambda) = 0,\quad  & \forall\:\delta\lambda \in\mathcal{M},\\
	\label{eq:weak:fluid_one} G_{\vec{u}}^\text{f}(p; \delta p) = 0,\quad & \forall\:\delta p \in\mathcal{P}_0.
	\end{alignat}
\end{subequations}
Note that $\vec{u}$ is still required as an input to compute the normal gap used in~(\ref{eq:weak:fluid_one}), thus we added the subscript ``$\vec{u}$'' for ${G}_{\vec{u}}^\text{f}$, however, for any given displacement field $\vec{u}$ the fluid-flow sub-problem becomes linear under the one-way coupling approach.

\section{Computational framework}

In this section, we discuss the solution of the formulated above coupled problem in a monolithic finite-element framework. Similarly to the previous section, we discuss handling of each sub-problem separately, and then we present the monolithic resolution algorithm which combines them all. We build the coupled framework on top of the standard displacement-based implicit finite element approach for elliptic problems of continuum mechanics, details of which may be found elsewhere, e.g. in~\citep{zienkiewicz1977finite}. Note that for brevity the same notations are preserved for discretized entities as were introduced in the continuous problem statement outlined above. Furthermore, in accordance with the problem statement presented in Section~\ref{sec:coupled_problem_statement} we use the small deformations and small rotations assumptions. Nevertheless, the necessary modifications to take into account large deformations and/or large rotations can be added into presented framework. The developed monolithic finite element framework was implemented in the finite-element suite Z-set~\citep{Besson1997,zset2019}.

\subsection{Mechanical contact}
\label{ssec:contact}

The solution of the considered coupled problem requires partitioning of the interface into contact, fluid-flow and, possibly multiple, trapped fluid zones, see~\eqref{eq:prob:domain}. In order to make the identification of the interface status consistent, it appears natural to associate each face $\Gamma_\text{el}$ of the surface $\Gamma$ with a contact element, i.e. adopt the ``face-to-rigid-surface'' discretization approach, rather than the ``node-to-rigid-surface'' technique according to which a contact element consists of a single node of the surface $\Gamma$, see~\citep{wriggers2006b,konyukhov2012computational,yastrebov2013b} for more details. Note also that according to our formulation, a contact element is associated with each face of the surface $\Gamma$ irrespective of the actual status of a given face; criteria for determining if a contact element is active or not will be discussed below.

In order to apply the augmented Lagrangian method (see discussion in Section~\ref{sec:var}), we append a scalar Lagrange multiplier to each node of the potential contact surface $\Gamma$. Thus, on each face $\Gamma_\text{el}$  we consider interpolation of the gap and Lagrange multiplier function as, respectively:
\begin{equation}
\label{eq:interp}
g(\xi,\eta) = \sum\limits_{i=1}^{n} N_i(\xi,\eta) \, g_{i}, \quad \lambda(\xi,\eta) = \sum\limits_{i=1}^{n} N_i(\xi,\eta) \,\lambda_i,
\end{equation}
where $n$ is the number of nodes adjacent to a face $\Gamma_\text{el}$, $N_i$ is the shape function associated with the node $i$, $\{\xi,\eta\}$ are local parametrization coordinates of a finite-element face, see Fig.~\ref{fig:sketch_contact}. Furthermore, $g_i = g_{0i} + \vec{u}_i\cdot\vec{\nu}$ is the gap value of the node $i$, where $g_{0i}$ is the initial gap and $\vec{u}_i$ is the displacement vector of the given node. Note that the same shape functions are used here for interpolation of geometric gap and Lagrange multipliers, however, it is not a necessary condition. We used bilinear shape functions associated with quadrilateral faces of the discretized surface, nonetheless, polynomials of higher order can be utilized, see~\citep{puso2008segment}. 

\begin{figure}[t]
	\centering
	\includegraphics[width=0.6\textwidth]{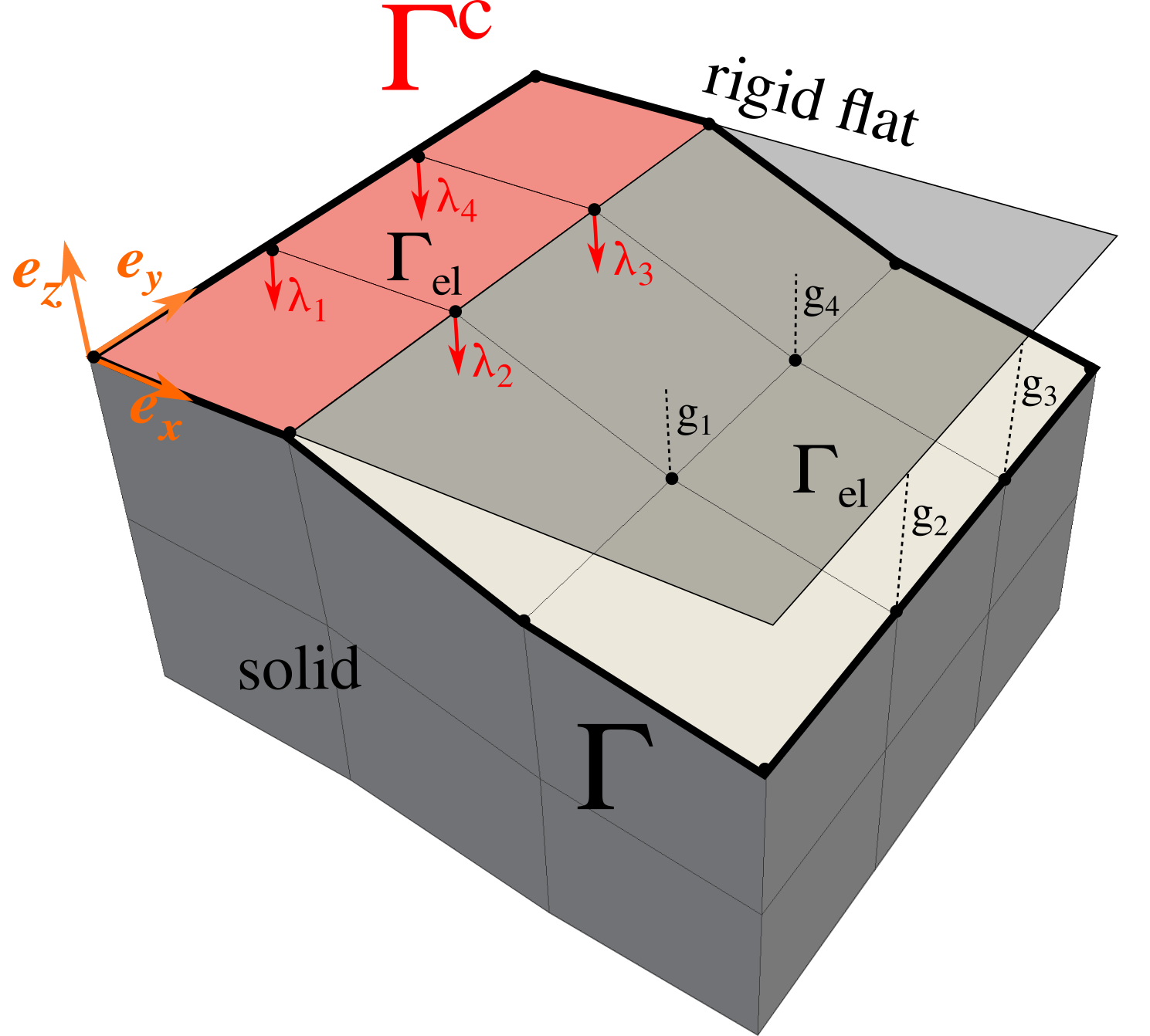} 
	\caption{Sketch of the interface highlighting contact elements: $\Gamma$ is the potential contact zone, $\Gamma^\text{c}$ is the active contact zone, $\Gamma_\text{el}$ is a face associated to one contact element, $\lambda$ is the Lagrange multiplier, which represents contact pressure and is attributed to each node of the  surface $\Gamma$, $g$ is the gap function: $g=0$ on $\Gamma^\text{c}$ and $g>0$ on $\Gamma\setminus\Gamma^\text{c}$.}
	\label{fig:sketch_contact}
\end{figure}

Due to the considered ``face-to-rigid-surface'' discretization and associated interpolation~\eqref{eq:interp}, the contact constraints~\eqref{eq:hsm} cannot be satisfied point-wise on the surface $\Gamma$.  
To overcome this inconsistency, we follow the mortar approach~\citep{puso2004_3d,puso2004mortar} and consider the third condition in~\eqref{eq:hsm} in the integral form over the surface $\Gamma$:
\begin{equation}
\label{eq:hsm_int}
\int\limits_{\Gamma} \lambda \, g \, d\Gamma = \sum\limits_{\text{el}} \:\int\limits_{\Gamma_\text{el}} \lambda \, g \, d\Gamma = 0,
\end{equation}
where the summation is performed over all contact elements on the surface $\Gamma$, and
the interpolated field of Lagrange multipliers $\lambda$ represents the normal traction $\sigma_n$. 
Substituting~\eqref{eq:interp} into \eqref{eq:hsm_int} and considering two first conditions of~\eqref{eq:hsm} at every node of the surface $\Gamma$, 
we obtain the following discrete (nodal) form of the contact conditions valid for each contact element, i.e. face $\Gamma_\text{el}$:
\begin{equation}
\label{eq:hsm_nodal}
\tilde{g}_i \geq 0, \; \lambda_i \leq 0, \; \lambda_i \, \tilde{g}_i = 0, \quad 1\leq i \leq n,
\end{equation}
where $\tilde{g}_i$ is termed as the integral (weighted) gap associated with the node $i$ of the face $\Gamma_\text{el}$ and is given by: 
\begin{equation}
\label{eq:int_gap}
\tilde{g}_i= \sum\limits_{j=1}^{n} \; g_j \int\limits_{\Gamma_\text{el}} N_i N_j \; d\Gamma = \sum\limits_{j=1}^{n} \; g_j \: I_{ij}.
\end{equation} 
In the finite-element framework weights $I_{ij}$ are calculated as:
\begin{equation}
\label{eq:mortar_coef}
I_{ij} = \int\limits_{-1}^{1} \int\limits_{-1}^{1} N_i \, N_j \: J \; d\xi \: d\eta,
\end{equation}
where $J$ is the Jacobian of the transformation of the physical coordinates $\vec{x} = [x, y, z]^\intercal$ on the surface $\Gamma_{\text{el}}$ to the face's coordinates in the parent space $\{\xi, \eta\}$:
\begin{equation}
\label{eq::jac}
J = \left|\frac{\partial\vec{x}}{\partial\xi}\times\frac{\partial\vec{x}}{\partial\eta}\right|, \quad \vec{x} =  \sum\limits_{i=1}^{n} \vec{x}_i N_i, 
\end{equation}
and $\vec{x}_i$ is the position of the $i$-th node of the face. Using Gauss quadrature rule, the integral in~\eqref{eq:mortar_coef} is computed as:
\begin{equation}
I_{ij} = \sum\limits_{k=1}^{n_\text{gp}} w_k \, N_i(\xi_k, \eta_k)N_j(\xi_k, \eta_k)J(\xi_k, \eta_k),
\end{equation}
where $n_\text{gp}$ is the number of Gauss points associated with the face $\Gamma_\text{el}$, $w_k$ is the weight coefficient of the $k$-th Gauss point, and $\{\xi_k, \eta_k\}$ are its coordinates in the parent space.

Next, in order to fulfil the contact constraints~\eqref{eq:hsm_nodal} on each element, we apply the augmented Lagrangian method and formulate the ``potential energy'' of a contact element as follows:
\begin{equation}
\label{eq:cont_energy}
W^\text{c}_\text{el} = \sum\limits_{i=1}^n
\begin{cases}
\displaystyle\lambda_i \,\tilde{g}_i + \frac{\epsilon}{2} \,\tilde{g}_i^2, \quad & \text{if}\;\hat{\lambda}_i \leq 0, \\
\displaystyle-\frac{1}{2\epsilon} \lambda_i^2,  \quad &  \text{if}\;\hat{\lambda}_i > 0,
\end{cases}
\end{equation}
which can be seen as the discretized version of the contact contribution to the augmented Lagrangian~\eqref{eq:aug_lag}. Note that $\epsilon$ acts here as the augmentation parameter with units [N/m$^5$].

The following notation of the augmented Lagrange multiplier is introduced: $\hat{\lambda}_i = \lambda_i + \epsilon \tilde{g}_i$, the sign of which defines the contact state of the node: if $\hat{\lambda}_i \leq 0$ the node is in the active state, while if $\hat{\lambda}_i > 0$ the node is not in contact. It is important to note that the augmented Lagrange multiplier depends on the weighted gap~\eqref{eq:int_gap}, integrated over a particular face, and therefore the state of each node is always evaluated with respect to a particular element, i.e. the same node can be in active state with respect to one adjacent element and in non-active state with respect to another. Furthermore, we term hereinafter an element as active if at least one of its nodes has $\hat{\lambda}_i\leq 0$ (with respect to this element), and inactive otherwise, which is important for a consistent partitioning of the interface into contact, fluid-flow and trapped fluid zones.

In order to find the contribution of each contact element to the balance of virtual work, we calculate the variation of~\eqref{eq:cont_energy}:
\begin{equation}
\label{eq:cont_work}
\delta W_{\text{el}}^\text{c} = \sum\limits_{i=1}^{n}
\begin{cases}
\displaystyle\hat{\lambda}_i\sum\limits_{j=1}^{n} I_{ij}\frac{\partial g_j}{\partial \vec{u}_{j}} \cdot \delta\vec{u}_{j} + \tilde{g}_i \delta\lambda_i, & \hat{\lambda}_i\leq 0 \\[6pt]
\displaystyle-\frac{1}{\epsilon} \lambda_i \delta \lambda_i, & \hat{\lambda}_i > 0
\end{cases}
\end{equation}
where $\vec{u}_j$ is the displacement vector of the node $j$. Note that in accordance with the infinitesimal strain formulation, integrals $I_{ij}$ are not variated. To linearize the problem, we calculate the second variation of the virtual work $\delta W^\text{c}_{\text{el}}$:
\begin{equation}
\label{eq:sec_var}
\displaystyle\Delta \delta W_{\text{el}}^\text{c} = \sum\limits_{i=1}^{n}
\begin{cases}
\displaystyle  \epsilon \sum\limits_{j=1}^{n} I_{ij} \frac{\partial g_j}{\partial \vec{u}_j} \cdot \delta \vec{u}_j \; \sum\limits_{k=1}^{n} I_{ik} \frac{\partial g_k}{\partial \vec{u}_k} \cdot \Delta \vec{u}_k +\sum\limits_{j=1}^{n} I_{ij} \frac{\partial g_j}{\partial \vec{u}_j} \cdot \left(\delta \vec{u}_j \Delta \lambda_i  + \Delta \vec{u}_j \delta \lambda_i \right)
\displaystyle, & \hat{\lambda}_i\leq 0 \\
\displaystyle -\frac{1}{\epsilon} \delta \lambda_i \Delta \lambda_i, & \hat{\lambda}_i > 0.
\end{cases}
\end{equation}

Finally, the virtual work~\eqref{eq:cont_work} and its variation~\eqref{eq:sec_var} could be expressed in a compact form, introducing the residual vector $\mathbf{R}^\text{c}$ and the tangent matrix $\mathbf{K}^\text{c}$ of a contact element:
\begin{equation} 
\label{eq:newton_c}
\delta W_{\text{el}}^\text{c} = {\begin{bmatrix}
		\mathbf{R}^\text{c}_{\vec{u}} \\[6pt]
		\mathbf{R}^\text{c}_{\lambda} 
		\end{bmatrix}}\tp \; \begin{bmatrix}
{\delta \vec{u}}  \\[6pt]
{\delta \lambda} 
\end{bmatrix}, \quad 
\Delta\delta W_{\text{el}}^\text{c} = \begin{bmatrix}
{\Delta \vec{u}}  \\[6pt]
{\Delta \lambda} 
\end{bmatrix}\tp \;
{\begin{bmatrix}
		\mathbf{K}^\text{c}_{\vec{u}\vec{u}} && \mathbf{K}^\text{c}_{\vec{u}\lambda} \\[6pt]
		\mathbf{K}^\text{c}_{\lambda\vec{u}} && \mathbf{K}^\text{c}_{\lambda\lambda} 
		\end{bmatrix}}\;  
\begin{bmatrix}
{\delta \vec{u}}  \\[6pt]
{\delta \lambda} 
\end{bmatrix},
\end{equation}
where for brevity we use the notations $\delta \vec{u} = [\delta \vec{u}_1, \ldots, \delta \vec{u}_n]\tp$, $\Delta \vec{u} = [\Delta \vec{u}_1, \ldots, \Delta \vec{u}_n]\tp$ and $\delta \lambda = [\delta \lambda_1, \ldots, \delta \lambda_n]\tp$, $\Delta \lambda = [\Delta \lambda_1, \ldots, \Delta \lambda_n]\tp$.  
Ready-to-implement expressions of the outlined components of $\mathbf{R}^\text{c}$ and $\mathbf{K}^\text{c}$ are given in Appendix~\ref{sec:app_contact}. Note that in the frictionless case considered here the tangent matrix of the contact element is symmetric, i.e. $\mathbf{K}^\text{c}_{\vec{u}\lambda} = \mathbf{K}^\text{c}_{\lambda\vec{u}}$.

\subsubsection{Post-processing computation of the real contact area\label{sec:contact_area}}

\begin{figure}[t]
	\centering
	\includegraphics[width=0.4\textwidth]{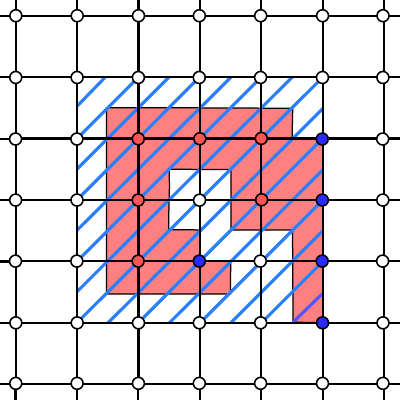} 
	\caption{
	Sketch on the computation of the real contact area. 
	Red circles represent nodes which are active with respect to all adjacent elements, 
	blue circles correspond to nodes which are active with respect to some of the adjacent elements but not all of them, and white circles are nodes not active with respect to any adjacent element. 
	The hatched blue area represents the contact area computed by summing up areas of all active elements (which have at least one active node), see~\eqref{eq:area_all}. The shaded red area is obtained by a refined approach of summing up for each element contributions to the contact area from each Gauss points nearest to an active node, see~\eqref{eq:area_gp}. }
	\label{fig:sketch_area}
\end{figure}

Consideration of a contact element as being active if at least one of its nodes is in active state is required for a consistent partitioning of the interface, which will be discussed in detail below. However, during the post-processing stage, different methods may be applied to compute the real contact area. 
A straightforward approach is to sum up areas $A_\text{el}$ of faces $\Gamma_\text{el}$ associated with active elements (hatched blue area in Fig.~\ref{fig:sketch_area}):
\begin{equation}
\label{eq:area_all}
A_\text{el} = \begin{cases}
\sum\limits_{k=1}^{n_\text{gp}} w_k \, J(\xi_k, \eta_k), & \exists\, i \in [1,\ldots,n] \text{ s.t. } \hat{\lambda}_i\leq 0 \\ 
0, & \forall \, i \in [1,\ldots,n] \quad \hat{\lambda}_i > 0.
\end{cases}
\end{equation}
However, our studies showed that this method of computation of the contact area leads to its significant overestimation. Therefore, we discuss here also a more precise approach for computing the real contact area: considering separately each contact element, add up to the contact area only contributions from Gauss points  closest to active nodes (shaded red area in Fig.~\ref{fig:sketch_area}):
\begin{equation}
\label{eq:area_gp}
A_\text{el} = \sum\limits_{i=1}^{n}\begin{cases}
w_i \, J(\xi_i, \eta_i), & \hat{\lambda}_i\leq 0 \\ 
0, & \hat{\lambda}_i > 0,
\end{cases}
\end{equation}
where $w_i$ and $(\xi_i , \eta_i)$ are the weight coefficient and coordinates of a Gauss point closest to the node $i$. Note that we assumed here that $n$ (the number of element's nodes) equals to $n_\text{gp}$ (the number of Gauss points of the corresponding face). However, if shape functions of a higher order than in~\eqref{eq:interp} are used for interpolation of the geometry and of the contact pressure, as was done, for example, in ~\citep{puso2008segment}, then a different refined approach to the real contact area computation tailored for such case may be required. The comparison of two discussed approaches to real contact area computation will be presented in Section~\ref{sec:contactareacomp}.

It is important to note, that the refined approach can be used when the number of Gauss points is not equal to the number of nodes of an element if the status is checked at Gauss points rather than at nodes. However, to make such a computation of the real contact area consistent with the augmented Lagrangian method used to solve the contact problem, the formulation presented in~\eqref{eq:cont_energy}-\eqref{eq:sec_var} should be slightly modified. In particular, the criterion for computing the element's contribution to the residual vector and the tangent matrix, i.e. the value of the augmented Lagrange multiplier, should also be checked at Gauss points. Our preliminary study of such contact formulation showed results (in terms of the contact pressure and contact area) very similar to those of the refined approach presented in this work. At the same time, a formulation based on Gauss-point-wise criterion opens possibilities of extending the developed framework to the case of hierarchical basis functions, which is an interesting topic for future research.

\subsection{Thin fluid flow\label{ssec:flow}}

\begin{figure}[t]
	\centering
	\includegraphics[width=0.6\textwidth]{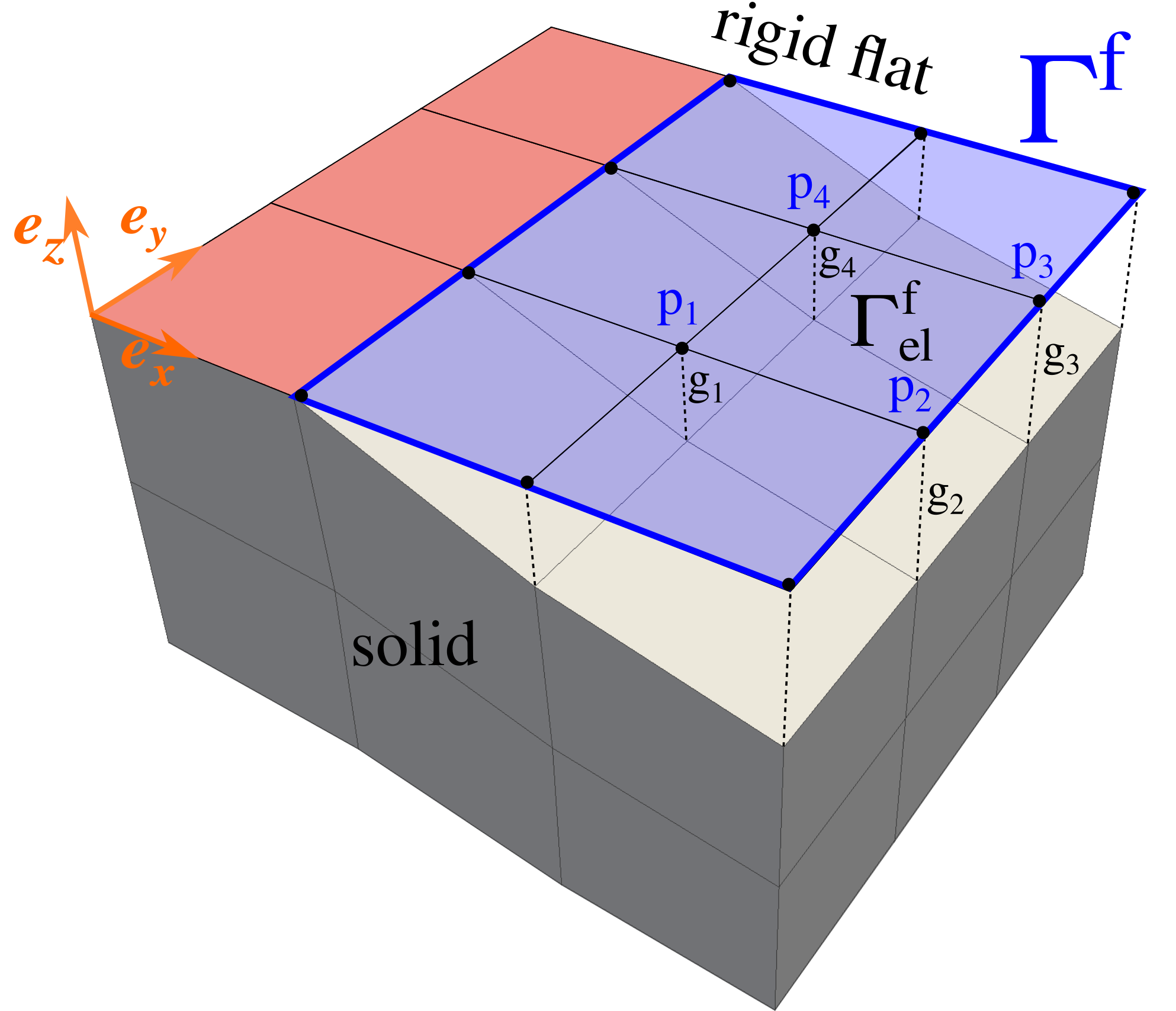} 
	\caption{ Sketch of the interface highlighting fluid flow elements: $\Gamma^\text{f}$ is the lubrication surface (subset of plane $z=0$), where the Reynolds equation is defined, $\Gamma^\text{f}_\text{el}$ is a face on this surface associated to one fluid flow element, $p_i$ is the fluid pressure DOF, added to each node of the surface of the solid (note that the constant fluid pressure across the film thickness is assumed).}
	\label{fig:sketch_fluid}
\end{figure}

Here we discuss the implementation of the weak form of the fluid flow problem~\eqref{eq:weak:fluid} into the finite-element framework. 
In order to concentrate the reader's attention on the aspects of the coupling, we consider only trivial Neumann boundary conditions on a portion of the fluid boundary, i.e. the prescribed zero fluid flux $q_0 = 0$ on $\gamma^q\subset\partial\Gamma^\text{f}$, however, a non-trivial Neumann boundary condition could be included in the present formulation in the standard manner.

Following the assumption of the constant fluid pressure across the film thickness, we attribute a fluid pressure DOF to each node of the surface $\Gamma$  and define a finite element for the fluid transport problem for each face of the surface $\Gamma^\text{f}$, formed by faces of $\Gamma^\text{fsi}$ projected onto the rigid flat, see Fig.~\ref{fig:sketch_fluid}. We use the same interpolation for the gap as in the contact problem, see~\eqref{eq:interp}, and the same interpolation is used for the fluid pressure $p$ and the test function $\delta p$:
\begin{equation}
\label{eq:shape_fl}
p = \sum\limits_{i=1}^n N_i(\xi,\eta)\,p_i, \quad \delta p = \sum\limits_{i=1}^n N_i(\xi,\eta)\,\delta p_i,
\end{equation}
where $n$ is the number of nodes which belong to a given element. Substituting these expressions into~\eqref{eq:coup:fluid} we obtain the following contribution of one fluid-flow element to the balance of the virtual work:
\begin{equation}
\label{eq:fluid:first_var}
\delta W^\text{f}_\text{el} =  \sum\limits_{i=1}^n\delta p_{i}\int\limits_{-1}^1 \int\limits_{-1}^1 \left(\sum\limits_{k=1}^n N_k \: g_k\right)^3 \left( \mathbf J^{-1} \sum\limits_{j=1}^n \nabla N_j p_j \right) \left(\mathbf J^{-1} \nabla N_i\right) \det (\mathbf{J}) \: d\xi d\eta,
\end{equation}
where $(\xi,\eta)$ are coordinates in the parent space, $\mathbf{J}$ is the Jacobian matrix defined as:
\begin{equation}
\label{eq:jac_f}
\mathbf{J} = \begin{bmatrix}
\frac{\partial x}{\partial \xi} && \frac{\partial x}{\partial\eta} \\[6pt]
\frac{\partial y}{\partial \xi} && \frac{\partial y}{\partial\eta}
\end{bmatrix},
\end{equation}
and $\det(\mathbf{J})$ is its determinant. The second variation reads:
\begin{align}
\label{eq:work_fl}
\Delta \delta W^\text{f}_\text{el} &= \sum\limits_{i=1}^n\delta p_{i}\Bigg\{\sum\limits_{j=1}^n\Delta p_{j}\int\limits_{-1}^1 \int\limits_{-1}^1 \left(\sum\limits_{k=1}^n N_k \: g_k\right)^3 \left(\mathbf{J}^{-1} \nabla N_j\right) \left(\mathbf{J}^{-1} \nabla N_i \right) \det(\mathbf{J}) \: d\xi d\eta\nonumber\\	
&+\sum\limits_{l=1}^n\frac{\partial g_{l}}{\partial \vec{u}_l}\Delta \vec{u}_l\int\limits_{-1}^1\int\limits_{-1}^1 3 \left(\sum\limits_{k=1}^n N_k \: g_k\right)^2 N_l \left(\mathbf{J}^{-1} \sum\limits_{j=1}^n \nabla N_j p_{j}\right) \left(\mathbf{J^{-1}} \nabla N_i \right) \det(\mathbf{J}) \; d\xi d\eta \Bigg\}.
\end{align}
Note that the variation of the Jacobian matrix $\mathbf{J}$, components of which are computed on the projection of faces of the solid's surface on the rigid flat, is not considered due to assumptions of small deformations and small rotations. Introducing the residual vector $\mathbf{R}^\text{f}$ and the tangent matrix $\mathbf{K}^\text{f}$ of a fluid flow element (explicitly given in Appendix~\ref{sec:app_fluid_flow}), we may write:
\begin{equation}
\label{eq:newton_f} 
\delta W^\text{f}_\text{el} = \begin{bmatrix}
\mathbf{R}^\text{f}_{p} \\[6pt]
0 
\end{bmatrix}\tp \; \begin{bmatrix}
{\delta p}  \\[6pt]
{\delta \vec{u}} 
\end{bmatrix}, \quad 
\Delta \delta W^\text{f}_\text{el} = \begin{bmatrix}
{\Delta p}  \\[6pt]
{\Delta \vec{u}} 
\end{bmatrix}\tp \;
\begin{bmatrix}
\mathbf{K}^\text{f}_{pp} && 0 \\[6pt]
\mathbf{K}^\text{f}_{\vec{u}p} && 0
\end{bmatrix}\;  \begin{bmatrix}
{\delta p}  \\[6pt]
{\delta \vec{u}} 
\end{bmatrix},
\end{equation}
with $\delta p = [\delta p_1, \ldots, \delta p_n]\tp$ and similarly $\Delta p = [\Delta p_1, \ldots, \Delta p_n]\tp$, whereas $\delta \vec u$ and $\Delta \vec u$ were given in the previous section. Note, that the presented formulation of the fluid-flow element was derived for the two-way coupled problem, however, it is also suitable for the one-way coupling. The only required modification is the assumption of the rigid walls of the solid, according to which the variation of the virtual work with respect to the displacement in~\eqref{eq:work_fl} vanishes and consequently $\mathbf{K}^\text{f}_{\vec{u}p} = 0$.

\subsection{Fluid-structure interface\label{ssec:fsi}}

To include the virtual work of fluid tractions on the solid's surface~\eqref{eq:coup:solid:fsi} into the numerical framework, we associate a fluid-structure interface element with each face of the surface $\Gamma^\text{fsi}$, see Fig.~\ref{fig:sketch_fsi}. We use the same interpolation of the fluid pressure and the gap, as in~\eqref{eq:shape_fl} and~\eqref{eq:interp}, respectively.
\begin{figure}[t]
	\centering
	\includegraphics[width=0.6\textwidth]{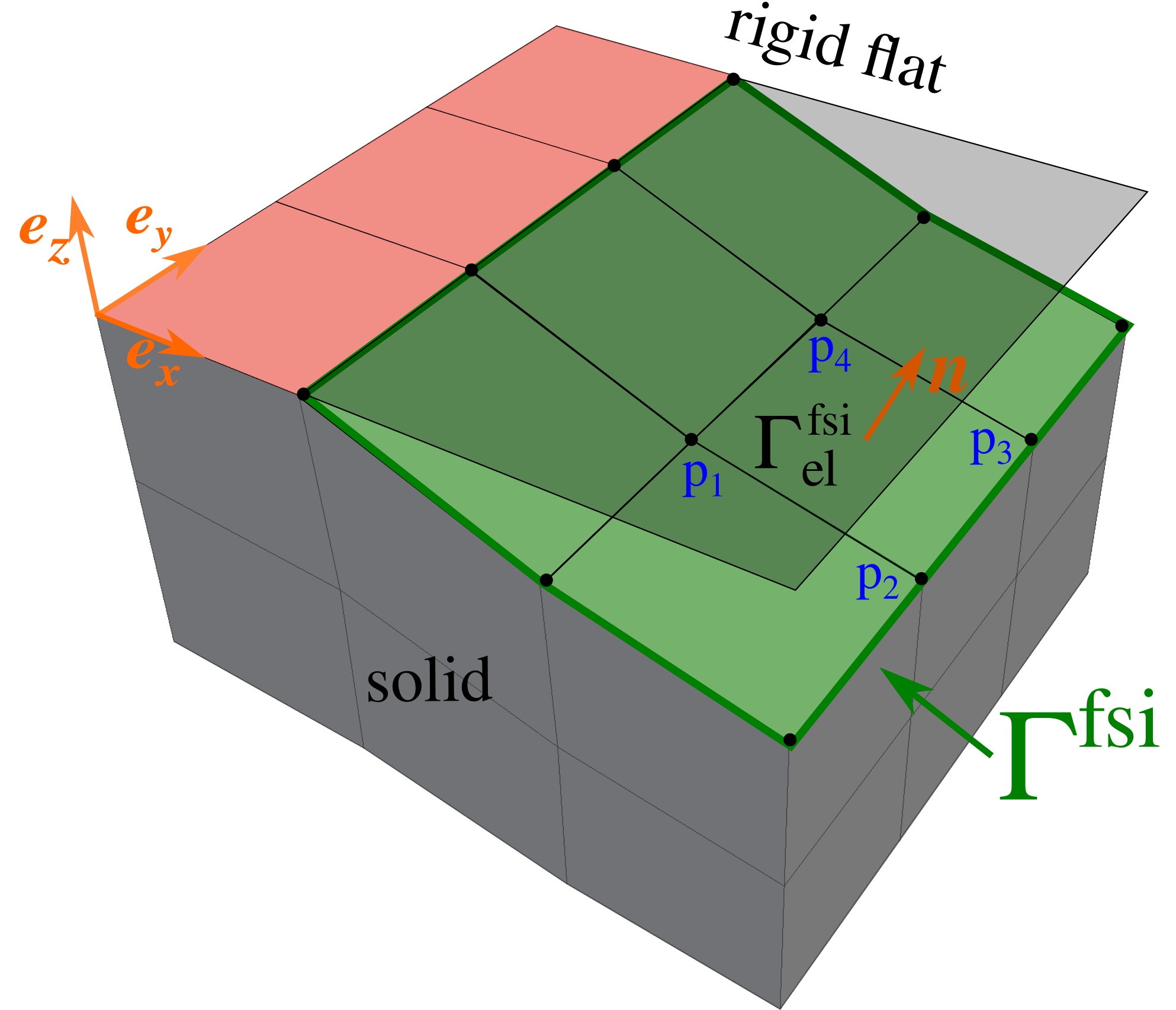} 
	\caption{Sketch of the interface highlighting FSI elements: $\Gamma^\text{fsi}$ is fluid-structure interface on the surface of the solid, $\Gamma^\text{fsi}_\text{el}$ is a face associated to one FSI element, $p$ is the fluid pressure at each node of this surface and $\vec{n}$ is the outer normal.}
	\label{fig:sketch_fsi}
\end{figure}
Therefore, the contribution of each fluid-structure interface element to the balance of virtual work reads:
\begin{align}
\delta W^\text{fsi}_\text{el} &= \sum\limits_{i=1}^n \delta \vec{u}_i \Bigg\{\sum\limits_{j=1}^n p_j \int\limits_{-1}^1 \int\limits_{-1}^1 \vec{n} \, N_i N_j \, J \, d\xi d\eta+\frac{1}{2}\sum\limits_{k=1}^n g_k \int\limits_{-1}^1 \int\limits_{-1}^1 \left(\mathbf{J}^{-1} \sum\limits_{l=1}^n \nabla N_l p_{l}\right) N_i N_k \, J \, d\xi d\eta \Bigg\},
\end{align}
where the gradients of shape functions $\nabla N_l$ are computed on the projection of the face $\Gamma^\text{fsi}$ on the rigid flat, i.e. in the same sense as in Section~\ref{ssec:flow}. The corresponding Jacobian matrix $\mathbf{J}$ was defined in~\eqref{eq:jac_f}, and the normal $\vec{n}$ is given by:
\begin{equation}
\label{eq:normal}
\displaystyle\vec{n} = \frac{\frac{\partial\vec{x}}{\partial\xi}\times\frac{\partial\vec{x}}{\partial\eta}}{\left|\frac{\partial\vec{x}}{\partial\xi}\times\frac{\partial\vec{x}}{\partial\eta}\right|}, \quad \vec{x} =  \sum\limits_{i=1}^{n} N_i(\xi, \eta) \, \vec{x}_i.
\end{equation}
Note that here the Jacobian $J$ is not computed on the lubrication surface and thus is not given by $\det(\mathbf{J})$, but it is computed on the surface of the deformable solid as in~\eqref{eq::jac}. The second variation then takes the following form:
\begin{align}
\Delta \delta W^\text{fsi}_\text{el} = \sum\limits_{i=1}^n \delta \vec{u}_i \Bigg\{&\sum\limits_{j=1}^n \Delta p_j \int\limits_{-1}^1 \int\limits_{-1}^1 \vec{n} \, N_i N_j \, J \, d\xi d\eta+\frac{1}{2}\sum\limits_{k=1}^n \frac{\partial g_k}{\partial \vec{u}_k}\Delta\vec{u}_k \int\limits_{-1}^1 \int\limits_{-1}^1 \left(\mathbf{J}^{-1} \sum\limits_{l=1}^n \nabla N_l p_{l}\right) N_i N_k \, J \, d\xi d\eta \nonumber\\
&+\frac{1}{2}\sum\limits_{k=1}^n g_k \sum\limits_{l=1}^n \Delta p_l \int\limits_{-1}^1 \int\limits_{-1}^1 \left(\mathbf{J}^{-1} \nabla N_l \right) N_i N_k \, J \, d\xi d\eta\Bigg\},
\end{align}
where variations of the Jacobian $J$, the matrix $\mathbf{J}$ and the normal vector $\vec{n}$ are not considered due to assumptions of small deformations and small rotations.

Finally, the associated virtual work and its variation could be expressed in a compact form using the residual vector $\mathbf{R}^\text{fsi}$ and the tangent matrix $\mathbf{K}^\text{fsi}$ of a FSI element (details can be found in Appendix~\ref{sec:app_fsi}):
\begin{equation} 
\label{eq:newton_fsi}
\delta W^\text{fsi}_\text{el} = \begin{bmatrix}
\mathbf{R}_{\vec{u}}^\text{fsi} \\[6pt]
0 
\end{bmatrix}\tp \; \begin{bmatrix}
{\delta \vec{u}}  \\[6pt]
{\delta p}
\end{bmatrix}, \quad 
\Delta\delta W^\text{fsi}_\text{el} = \begin{bmatrix}
{\Delta \vec{u}}   \\[6pt]
{\Delta p} 
\end{bmatrix}\tp \;
\begin{bmatrix}
\mathbf{K}_{\vec{u}\vec{u}}^\text{fsi} &&  0 \\[6pt]
\mathbf{K}_{p\vec{u}}^\text{fsi}  && 0
\end{bmatrix}\;  \begin{bmatrix}
{\delta \vec{u}} \\[6pt]
{\delta p}
\end{bmatrix}.
\end{equation}
Note that in the case of one-way coupling, the action of the fluid pressure on the surface of the solid is neglected, so that the virtual work~\eqref{eq:coup:solid:fsi} vanishes and no contribution of the FSI element is included into the global system, i.e. $\mathbf{R}_{\vec{u}}^\text{fsi} = 0, \; \mathbf{K}_{\vec{u}\vec{u}}^\text{fsi} = 0, \mathbf{K}_{p\vec{u}}^\text{fsi} = 0.$ 

\subsection{Trapped fluid zones}
\label{ssec:tf}
In order to take into account the effect of pressurized volumes of trapped fluid~\eqref{eq:coup:solid:trap} on the coupled problem, we follow~\citep{shvarts2018trapped} and use the nonlinear penalty method to simulate the behaviour of compressible fluid with a pressure-dependent bulk modulus. We discuss here two possible approaches for implementing this model into the finite-element framework. First, we present a ``superelement'' formulation for a trapped-fluid element containing all faces of one trapped fluid zone $\Gamma^\ttf_i, i=\overline{1, n_\text{tf}}$.  A possible standard finite-element formulation which computes contributions from each face of the trapped fluid zone separately is also stated; benefits and drawbacks of these two approaches are briefly discussed.

\subsubsection{Superelement formulation}

In the finite-element framework the volume of the $i$-th trapped zone~\eqref{eq::coup:gap_vol} can be calculated by the following formula:
\begin{equation}
\label{eq:tf_gap_vol}
V^\text{tf}_i(\mathbf{U}) = \sum\limits_{k=1}^{m_i}\:\int\limits_{-1}^1 \int\limits_{-1}^1 \sum\limits_{j=1}^{n_k} N_j \, g(\vec{u}_{j})\, (-\vec{n}_k\cdot\vec{\nu})\,J \, d\xi d\eta,
\end{equation}
where the summation with respect to index $k=\overline{1,m_i}$ is performed over all faces of the surface $\Gamma^\ttf_i$, and the summation with respect to $j=\overline{1, n_k}$ is over all nodes of the $k$-th face. Thus, we denote by $\mathbf{U} = [\vec{u}_1, \ldots \vec{u}_j, \ldots\vec{u}_M]\tp$ vector of displacements of all $M$ nodes on the surface $\Gamma^\ttf_i$, which shall serve as the DOF vector for the trapped fluid superelement. Next, $g(\vec{u}_{j})$ is the normal gap computed for the $j$-th node of the $k$-th face and $N_j(\xi,\eta)$ is the shape function associated with this node;  $J$ is the Jacobian defined in~\eqref{eq::jac}. Finally, $\vec{n}_k$ is the normal to the $k$-th face, which can be computed in the same way as in~\eqref{eq:normal}.
The integral in~\eqref{eq:tf_gap_vol} can be calculated using the Gauss quadrature rule as:
\begin{equation}
V^\text{tf}_i(\mathbf{U}) = \sum\limits_{k=1}^{m_i} \sum\limits_{l=1}^{n_k^\text{gp}} \: \sum\limits_{j=1}^{n_k}g(\vec{u}_{j})\,N_j(\xi_l, \eta_l)\, (-\vec{n}_k(\xi_l, \eta_l)\cdot\vec{\nu})\,J(\xi_l, \eta_l)\, w_l,
\end{equation}
where $n_k^\text{gp}$ is the number of Gauss points associated with the $k$-th face of the surface $\Gamma^\ttf_i$, $w_l$ is the weight coefficient of the $l$-th Gauss point, $(\xi_l, \eta_l)$ are its coordinates in the parent space. 

Therefore, using the expression for the virtual work~\eqref{eq:weak:trap}, we may write the residual vector and the tangent matrix for the trapped fluid superelement as:
\begin{align}
\label{eq::elem_tf_nonlin_penalty}
\textbf{R}^{\ttf_i}_{\vec{u}} =
-\left\{\left(\frac{K_0}{K_1} + p^{\ttf}_{i0}\right) \left(\frac{V^\text{tf}_i(\mathbf{U})}{V^\ttf_{0}}\right)^{-K_1} - \frac{K_0}{K_1}\right\}
\left[\frac{\partial V^\text{tf}_i(\mathbf{U})}{\partial \mathbf{U}}\right],\nonumber\\
\textbf{K}^{\ttf_i}_{\vec{u}\vec{u}} =
\left(\frac{K_0}{K_1} + p^{\ttf}_{i0}\right) \frac{K_1}{V^\ttf_{0}} \left(\frac{V^\text{tf}_i(\mathbf{U})}{V^\ttf_{0}}\right)^{-K_1-1}
\left[\frac{\partial V^\text{tf}_i(\mathbf{U})}{\partial \mathbf{U}}\right] \otimes \left[\frac{\partial V^\text{tf}_i(\mathbf{U})}{\partial \mathbf{U}}\right]\tp,
\end{align}
where $\otimes$ is tensor product. Note that we used the small strain formulation and neglected the variation of normals, and therefore the matrix of second derivatives $\partial^2 V^\ttf_i/\partial\mathbf{U}^2$ is zero, which simplifies the formulation of $\textbf{K}^{\text{tf}i}_{\vec{u}\vec{u}}$. 

\subsubsection{Standard finite-element formulation}

Alternatively to the superelement formulation presented above, a standard finite-element formulation can be used, according to which the residual vector $\mathbf{R}^{\ttf_i}_{\vec{u}}$ and the tangent matrix $\mathbf{K}^{\ttf_i}_{\vec{u}\vec{u}}$ are assembled using separate contributions from each face of the trapped fluid zone. However, in application to the considered problem the standard approach is bound to certain limitations. Indeed, the volume of the gap~\eqref{eq:tf_gap_vol} can be computed as sum of volumes $V_k$ corresponding to each single face:
\begin{equation}
\label{eq:vol:separ}
V^\ttf_i (\mathbf{U}) = \sum\limits_{k=1}^{m_i} V_k(\mathbf{U}_k),\quad V_k(\mathbf{U}_k) = \int\limits_{-1}^1 \int\limits_{-1}^1 \sum\limits_{j=1}^{n_k} N_j\, g(\vec{u}_{j})\, (-\vec{n}_k\cdot\vec{\nu})\,J \, d\xi d\eta,
\end{equation}
where $\mathbf{U}_k$ is the vector of displacements of nodes of the $k$-th face only.
However, the behaviour of fluid in each elemental volume $V_k$ cannot be considered to follow the same constitutive behaviour as the entire trapped fluid volume $V^\ttf_i$, i.e. the union of volumes $V_k$ should react as an ensemble.
Accordingly, the tangent matrix $\mathbf{K}^{\ttf_i}_{\vec{u}\vec{u}}$ which includes the tensor product $\left[\partial V^\ttf_i/\partial \mathbf{U}\right] \otimes \left[\partial V^\ttf_i/\partial \mathbf{U}\right]\tp$, see~\eqref{eq::elem_tf_nonlin_penalty}, is not sparse and cannot be constructed using the standard assembly process.

Nevertheless, the standard finite-element formulation can still be used to handle the trapped fluid model if the method of Lagrange multipliers and the penalty method are used simultaneously. In order to show that, following~\citep{abaqus2018}, we consider the contribution of the trapped fluid to the combined Lagrangian for the coupled problem as:
\begin{equation}
W^\ttf_i =-\lambda^\ttf_i \left[V^\ttf_i(\mathbf{U}) - V^{\ttf^*}_i(p^\ttf_i)\right],
\end{equation}
where $\lambda^\ttf_i$ is an additional Lagrange multiplier, $V^\ttf_i(\mathbf{U})$ is computed in the same sense as in~\eqref{eq::coup:gap_vol}, while $V^{\ttf^*}_i(p^\ttf_i)$ represents the volume of the trapped fluid as a function of its pressure, i.e. the inverse of the constitutive relation~\eqref{eq:coup:nonlin_comp}: 
\begin{equation}
\label{eq:inv_nonlin}
V^{\ttf^*}_i(p^\ttf_i) = V^\ttf_{i0}\left(\frac{p^\ttf_i + K_0/K_1}{p^\ttf_{i0} + K_0/K_1}\right)^{-1/K_1}.
\end{equation}
Realizing that the Lagrange multiplier $\lambda^\ttf_i$ is equivalent to the pressure in the $i$-th trapped fluid zone, we may 
substitute $p^\ttf_i$ by $\lambda^\ttf_i$ in~\eqref{eq:inv_nonlin} and express the variation of the term $W^\ttf_i$ as:
\begin{equation}
\delta W^\ttf_i = -\left[\lambda^\ttf_i\,\frac{\partial V^\ttf_i(\mathbf{U})}{\partial\mathbf{U}} \cdot\delta \mathbf{U} + \left(V^\ttf_i(\mathbf{U}) - V^{\ttf^*}_i(\lambda^\ttf_i)\right)\delta\lambda^\ttf_i - \lambda^\ttf_i\,\frac{\partial V^{\ttf^*}_i(\lambda^\ttf_i)}{\partial\lambda^\ttf_i}\,\delta\lambda^\ttf_i\right],
\end{equation}
which permits now to avoid the tensor product in the tangent matrix  $\mathbf{K}^{\ttf_i}_{\vec{u}\vec{u}}$ and apply the standard finite-element assembly, combining contribution from each face separately. 

Furthermore, if the problem under consideration involves multiple trapped fluid zones, then an additional Lagrange multiplier $\lambda^\ttf_i$ needs to be considered for each one of them. However, the number and the extent of trapped fluid zones can vary not only between load steps, but also between iterations of the Newton-Raphson method. The associated inevitable change of the size of the global DOF vector, and consequently, the global matrix, is undesirable, particularly for an implicit finite-element code. Moreover, it can make the algorithm for tracking trapped fluid zones (discussed below) more complex. 
Therefore, in our implementation we followed the proposed above approach of superelements for each trapped fluid zone, which does not require additional DOFs. 
Nevertheless, it is important to bear in mind that in this case the tangent matrix of the trapped fluid element is not sparse, which may increase considerably the storage space required for the construction of the global matrix, as well as its bandwidth.

\subsection{Monolithic coupling}

Following the monolithic approach to the coupled problem, we combine contributions of all discussed above sub-problems	 together, outlining the general structure of the global tangent matrix and residual vector, constructed for each iteration of the Newton-Raphson method. Moreover, we propose a novel algorithm for identification of the local status of the interface in the sense of partition given by Eq.~\eqref{eq:prob:domain}.

\subsubsection{The global residual vector and tangent matrix for the coupled problem}

 We will denote here by $\vec{v} = [\vec{u}, \lambda, p]^\intercal$ the global vector of nodal DOFs, consisting of displacement components, Lagrange multipliers for the contact problem, and fluid pressure values for the fluid flow problem, respectively. Note that the trapped fluid elements do not require any additional degrees of freedom, since the penalty method and the superelement formulation were used. Furthermore, by $\Delta\vec{v} = [\Delta\vec{u}, \Delta\lambda, \Delta p]^\intercal$ we shall denote the increment of the DOF vector, corresponding to a single iteration. Then the global system of equations takes the following form:
\begin{equation}
\label{eq:tangent_matrix}
\begin{bmatrix}
\mathbf{K}^{*}_{\vec{u}\vec{u}} && \mathbf{K}^\text{c}_{\vec{u}\lambda} && \mathbf{K}^\text{f}_{\vec{u}p} \\[6pt]
\mathbf{K}^\text{c}_{\lambda\vec{u}} && \mathbf{K}^\text{c}_{\lambda\lambda} && 0 \\[6pt]
\mathbf{K}^\text{fsi}_{p\vec{u}} && 0 && \mathbf{K}^\text{f}_{pp} 
\end{bmatrix} \;
\begin{bmatrix}
{\Delta\vec{u} } \\[6pt]
{\Delta\lambda} \\[6pt]
{\Delta p} 
\end{bmatrix}\;=
\;
-\begin{bmatrix}
\mathbf{R}^{*}_{\vec{u}}\\[6pt]
\mathbf{R}^\text{c}_{\lambda} \\[6pt]
\mathbf{R}^\text{f}_{p}  
\end{bmatrix},
\end{equation}
The matrix $\mathbf{K}^{*}_{\vec{u}\vec{u}}$ and vector $\mathbf{R}^{*}_{\vec{u}}$ are assembled using corresponding entities of all aforementioned sub-problems, introduced in~\eqref{eq:newton_c},~\eqref{eq:newton_f},~\eqref{eq:newton_fsi} and~\eqref{eq::elem_tf_nonlin_penalty}:
\begin{equation}
\label{eq:glob_mat}
\mathbf{K}^{*}_{\vec{u}\vec{u}} = \mathbf{K}^\text{s}_{\vec{u}\vec{u}} + \mathbf{K}^\text{c}_{\vec{u}\vec{u}} + \mathbf{K}^\text{fsi}_{\vec{u}\vec{u}} + \sum\limits_{i=1}^{n_\text{tf}}\mathbf{K}^{\text{tf}_i}_{\vec{u}\vec{u}},\quad
\mathbf{R}^{*}_{\vec{u}} = \mathbf{R}^\text{s}_{\vec{u}} + \mathbf{R}^\text{c}_{\vec{u}} + \mathbf{R}^\text{fsi}_{\vec{u}} +  \sum\limits_{i=1}^{n_\text{tf}}\mathbf{R}^{\text{tf}_i}_{\vec{u}},
\end{equation}
where $\vec{R}^\text{s}_{\vec{u}}$ and $\vec{K}^\text{s}_{\vec{u}\vec{u}}$ are the residual vector and tangent matrix of the solid mechanics problem in absence of contact constraints and fluid flow, computed in a standard way as:
\begin{equation}
\vec{R}^\text{s}_{\vec{u}} = \frac{\partial{\Pi^\text{s}(\vec{u})}}{\partial\vec{u}},\quad \vec{K}^\text{s}_{\vec{u}\vec{u}} = \frac{\partial^2{\Pi^\text{s}(\vec{u})}}{\partial\vec{u}^2}.
\end{equation}
If one prefers to perform simulation of the two-way coupling neglecting the presence of trapped fluid and considering only the effect of fluid pressure in the flow on the deformation of the solid, then contributions of all trapped fluid zones may be omitted in Eqs.~\eqref{eq:tangent_matrix}-\eqref{eq:glob_mat}. It is also important to note  that in case of two-way coupling the global matrix defined in~\eqref{eq:tangent_matrix} is not symmetric, since non-diagonal block terms $\mathbf{K}^\text{f}_{\vec{u}p}$ and $\mathbf{K}^\text{fsi}_{p\vec{u}}$ are obtained upon discretization of different equations, see~\eqref{eq:newton_f} and~\eqref{eq:newton_fsi}, and therefore are not equal in the general case. However, if the one-way coupling is considered, these terms vanish, rendering the global tangent matrix symmetric. To solve the system~\eqref{eq:glob_mat} in both one- and two-way coupling cases we use a direct multi-frontal solver MUMPS~\citep{amestoy2000multifrontal}  in a shared-memory parallel environment, while a distributed-memory version of the framework is a subject of future work.

Newton-Raphson iterations are performed until the norm of the global residual vector falls below a prescribed tolerance. However, for the coupled problem under study in order to ensure the balance between different fields, we consider separately the norms of the subsets of the residual vector corresponding to different types of DOFs: $\mathbf{R}^{*}_{\vec{u}}$,  $\mathbf{R}^\text{c}_{\lambda}$, $\mathbf{R}^\text{f}_{p}$, see~\eqref{eq:tangent_matrix}. Therefore, iterations are performed until all of the following conditions are simultaneously fulfilled:
\begin{equation}
\label{eq:newton_cond}
\frac{||\mathbf{R}^{*}_{\vec{u}}||_{2}}{||\mathbf{R}^\text{ext}_{\vec{u}}||_{2}} < \epsilon_{\vec{u}}, \quad ||\mathbf{R}^\text{c}_{\lambda}||_{\infty} < \epsilon_{\lambda}, \quad ||\mathbf{R}^\text{f}_{p}||_{\infty} < \epsilon_p,
\end{equation}
where $\epsilon_{\vec{u}}, \epsilon_{\lambda}, \epsilon_p$ are the error tolerance thresholds, chosen separately for each type of the DOF. 
Note that for the contact mechanics sub-problem a nodal vector of external loads $\mathbf{R}^\text{ext}$ is available, and, therefore, for the displacements residual $\mathbf{R}^{*}_{\vec{u}}$ we consider the \textit{relative} error, while for two other residuals we use the \textit{absolute} error criterion. The following notations are used for two different definitions of a norm of a vector $\mathbf{R}$:
\begin{equation}
||\mathbf{R}||_2 = \sqrt{\sum\limits_{i=1}^{N} R_i^2}, \quad||\mathbf{R}||_\infty = \max\limits_{i=1\ldots N} |R_i|,
\end{equation}
where $N$ is the length of this vector. Note that~\eqref{eq:newton_cond} renders the choice of the error thresholds $\epsilon_{\lambda}$ and $\epsilon_p$ problem-dependent, unless an appropriate scaling of the residual vector and the tangent matrix is performed.

\subsubsection{Resolution algorithm}

\begin{algorithm}[p]
	\caption{Resolution procedure for the coupled problem }
	\label{alg:newton}
	\begin{algorithmic}[1]
		\Require $v^i$, $i = 0$
		\Procedure{NewtonRaphsonLoop}{\hspace{0pt}}
		\Repeat 
		\ForAll{faces of $\Gamma$}
		\If{corresponding contact element is active}
		\State face's label $\gets$ \texttt{CONTACT}
		\Else 
		\State face's label $\gets$ \texttt{NONE}
		\EndIf
		\EndFor
		\ForAll{faces with nodes from inlet or outlet} \Comment{Dirichlet B.C.}
		\State \Call{DepthFirstSearch}{face, \texttt{FLOW}} 
		\EndFor		
		\ForAll{faces of $\Gamma$} 
		\State Construct $\mathbf{R}^\text{c}$ and $\mathbf{K}^\text{c}$ for corresponding contact element using~\eqref{eq:newton_c} 
		\Statex \hspace{35pt} \Comment{ALM permits to have contact elements in non-active zone} 
		\If{face's label $=$ \texttt{FLOW}} 		
		\State Construct $\mathbf{R}^\text{f}$ and $\mathbf{K}^\text{f}$ for corresponding fluid-flow element using~\eqref{eq:newton_f} 
		\If{two-way coupling}
		\State Construct $\mathbf{R}^\text{fsi}$ and $\mathbf{K}^\text{fsi}$ for corresponding FSI element using~\eqref{eq:newton_fsi} 
		\EndIf
		\EndIf
		\EndFor
		\If{two-way coupling}
		\State \Call{IdentifyTrappedZones}{\hspace{0pt}} 
		\ForAll{trapped fluid elements} 
		\State Construct $\mathbf{R}^\ttf$ and $\mathbf{K}^\ttf$ for trapped-fluid element using~\eqref{eq::elem_tf_nonlin_penalty}
		\EndFor
		\EndIf
		\State Solve system~\eqref{eq:tangent_matrix} for $\Delta \vec{v}^\text{i}$
		\State $\vec{v}^{i+1} \gets \vec{v}^\text{i} + \Delta\vec{v}^\text{i}$
		\State $i \gets i + 1$
		\Until{\eqref{eq:newton_cond} is validated}
		\EndProcedure
	\end{algorithmic}
\end{algorithm}

\begin{figure}[p]
	\centering
  \includegraphics[width=1.0\textwidth]{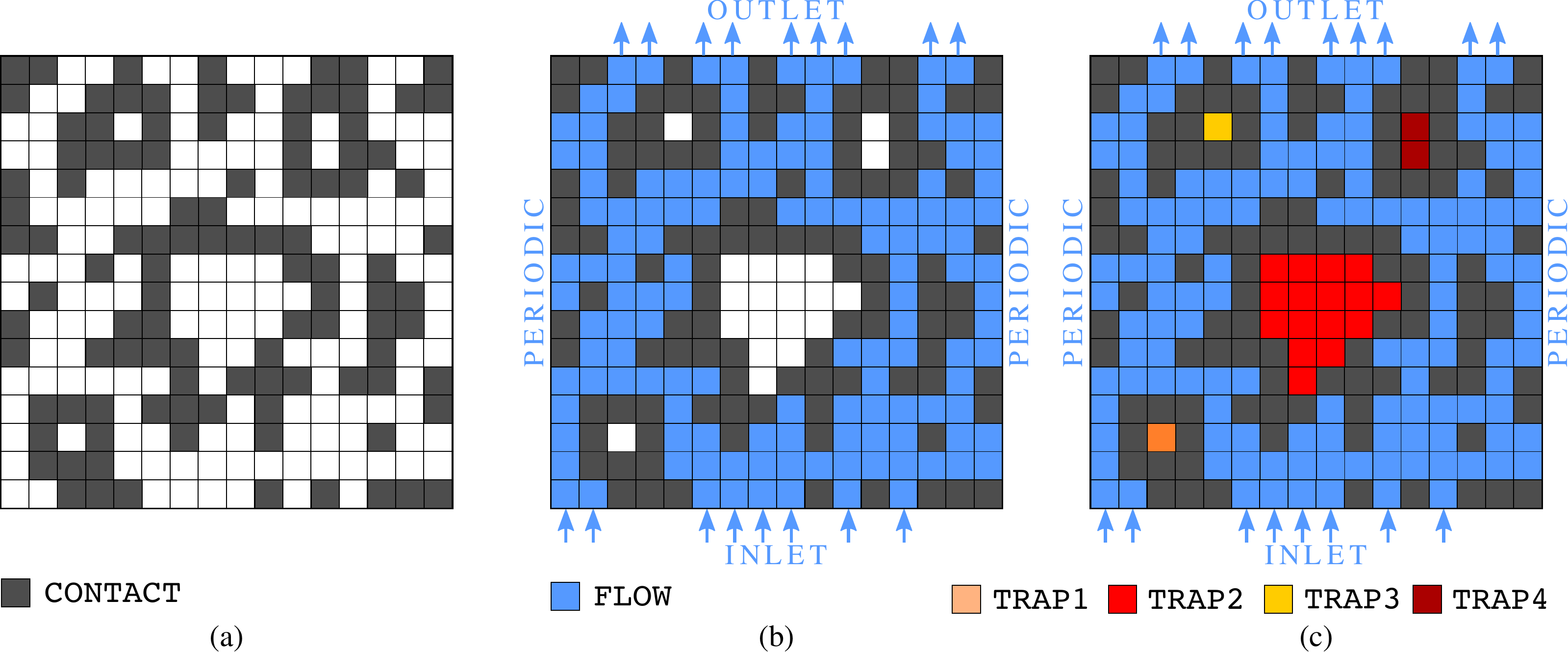} 
  \caption{Partitioning of the interface into active contact (a), fluid-flow (b) and trapped fluid zones (c), according to which every face of the surface $\Gamma$ is assigned a label from the following list: \{\texttt{CONTACT}, \texttt{FLOW}, \texttt{TRAP1}, \texttt{TRAP2}, \texttt{TRAP3}, \texttt{TRAP4}, ...\}. } 
  \label{fig:map_alg}
\end{figure}

\begin{algorithm}[p]
	\caption{Depth-first search (DFS)}
	\label{alg:dfs}
	\begin{algorithmic}[1]
		\Procedure{DepthFirstSearch}{face, \texttt{LABEL}}
		\If{face's label = \texttt{NONE}}
		\State face's label $\gets$ \texttt{LABEL}
		\ForAll{neighbours of face}
		\State \Call{DepthFirstSearch}{neighbour, \texttt{LABEL}}
		\EndFor
		\EndIf
		\EndProcedure
	\end{algorithmic}
\end{algorithm}

\begin{algorithm}[p]
	\caption{Trapped fluid zones identification procedure}
	\label{alg:trapped}
	\begin{algorithmic}[1]
		\Procedure{IdentifyTrappedZones}{\hspace{0pt}}
		\ForAll{trapped fluid elements} \Comment{the same list is used for all load steps}
		\If{element did not exist at the previous converged load step}
		\State delete trapped fluid element 
		\Else
		\State empty element's list of faces
		\EndIf
		\EndFor
		\ForAll{faces}
		\If{face's label = \texttt{NONE}}
		\If{face's label at the previous converged load step $=$ \texttt{TRAP+id}}
		\State id $\gets$ corresponding trapped fluid element's id
		\State \Call{DepthFirstSearch}{face, \texttt{TRAP+id}} 
		\Else
		\State Create new trapped fluid element
		\State id $\gets$ new number of trapped fluid elements
		\State \Call{DepthFirstSearch}{face, \texttt{TRAP+id}} 
		\EndIf
		\EndIf			
		\EndFor
		\ForAll{faces}
		\If{face's label = \texttt{TRAP+id}} 
		\State append face to corresponding element's list of faces
		\EndIf
		\EndFor
		\ForAll{trapped fluid elements}
		\If{element did not exist at the previous converged load step}
		\State Compute mean fluid pressure from the previous converged load step
		\Statex \hspace{20pt} \Comment{the initial pressure is the average fluid pressure over all faces of this element}
		\EndIf
		\EndFor
		\EndProcedure
	\end{algorithmic}
\end{algorithm}

Finally, a methodology for identification of local status (active contact, fluid flow, or trapped fluid) of interface elements remains to be defined.  Before starting the computation, we construct a graph of the interface, where each \textit{vertex} of the graph represents a face of the surface $\Gamma$. Two vertices are connected by a \textit{link} in the graph if the two corresponding faces of the surface $\Gamma$ share an edge, i.e. for a regular quadrilateral surface mesh we consider the so-called ``4-connected'' (von Neumann) neighbourhood.

At every iteration of the Newton-Raphson method we perform the following steps, summarized in~Algorithm~\ref{alg:newton}, which can be easily adjusted to one-way and two-way coupling approaches. We start by identifying active contact elements using the criterion presented in Subsection~\ref{ssec:contact}, i.e. a contact element is considered active if at least one of its nodes is active. Accordingly, each face which corresponds to an active contact element is assigned the label \texttt{CONTACT}, see Fig.~\ref{fig:map_alg}(a). Next, in order to locate the fluid-flow domain, we perform connected-component labelling of faces which are not in active contact using the depth-first search (DFS)~\citep{shapiro1996connected}, see Fig.~\ref{fig:map_alg}(b) and~Algorithm~\ref{alg:dfs}. Note that this recursive procedure is started from faces which feature fixed pressure due to Dirichlet boundary conditions at least in one node. Once every face in the fluid-flow domain is marked by the label \texttt{FLOW}, we continue the connected-component labelling of remaining non-contact faces to identify separately each trapped fluid zone (label \texttt{TRAP+id}, where \texttt{id} is the number of the trapped pool). The corresponding procedure is summarized in~Algorithm~\ref{alg:trapped}, see also Fig.~\ref{fig:map_alg}(c).

Since the behaviour of each trapped fluid pool depends on the volume of this pool at the moment of its formation, and also on the corresponding average pressure of the fluid inside, a modification of the standard connected-component labelling algorithm is necessary to track creation and evolution of trapped fluid zones. In particular, for each observed trapped fluid pool two cases are possible: some (or all) faces of this zone could have belonged to a trapped fluid zone identified at the end of the previous (converged) loading step, or the considered zone could correspond to a new trapped fluid pool formed at the current loading step. In the former case, the trapped fluid zone inherits the data (the initial volume and the fluid pressure) from the zone identified at the previous load step. In the latter case, the current volume of a newly created pool is stored and the initial pressure is computed as the mean of the fluid flow pressure values calculated at the end of the previous load step. 
Note that our study presented in Section~\ref{ssec:atoll} shows that the fluid flow pressure in a zone which would become trapped at the following load step is very close to being uniform (in case of reasonable change in load between steps). 

It is important to note, that in a simulation with an increasing external load, the splitting of a single trapped fluid pool into multiple zones is possible, 
which will not be recognized by the presented algorithm, i.e. these multiple pools will still be treated as one volume of trapped fluid. However, the effect of such a simplification on the transmissivity of rough contact interfaces, studied in Section~\ref{ssec:rough}, based on the observed results, is not expected to be significant. The opposite process, i.e. the merging of multiple trapped fluid zones into single one is also not covered by the presented algorithm. However, this process would require elimination of the contact area between separate pools, while the study of the trapped fluid problem presented in~\citep{shvarts2018trapped} shows that a considerable reduction of the contact area corresponds to a significantly higher external load than the one needed for the complete sealing of the interface. Nevertheless, aforementioned special cases can be included into the presented framework without considerable difficulties. At the same time, the opening of a trapped zone as described in~\citep{shvarts2018trapped} coupled to a fluid flow in the interface would necessarily result in a transient process and cannot be accurately taken into account in the quasi-static framework presented here.

Finally, at each iteration we compute the number of local status changes with respect to the previous iteration (or previous converged load step in case of the first iteration). 
At the $i$-th iteration this value is calculated as:
\begin{equation}
\label{eq:status_change}
S^i = \sum\limits_{j=1}^m \begin{cases}
1, & s^i_j \neq s^{i-1}_j \\
0, & s^i_j = s^{i-1}_j,
\end{cases}
\end{equation}
where $m$ is the total number of faces of the surface $\Gamma$, $s^i_j$ is the label of the $j$-th face, corresponding to the $i$-th iteration, and, accordingly, $s^{i-1}_j$ is the label of the same face at the previous iteration (or previous converged load step if $i=1$). The label, according to the Algorithm~\ref{alg:newton}, is from the list \{\texttt{CONTACT}, \texttt{FLOW}, \texttt{TRAP1}, \texttt{TRAP2}, \texttt{TRAP3}, ...\}. 
At the post-processing stage the number of local status changes permits to study the convergence of the Newton-Raphson method and verify the proposed resolution procedure.

\section{Examples and discussions}
\label{sex:examples}
In this section we present examples which show the capabilities of the proposed framework. We highlight the difference between the solutions obtained under one-way and two-way coupling approaches, and, in particular, we demonstrate the effect of the trapped fluid on the coupled problem. Additionally, we demonstrate the robustness of the resolution method showing the residual-wise and the status-wise convergence of the Newton-Raphson method, and also compare two different methods of the contact area computation.

\subsection{Fluid flow across a wavy channel with an ``atoll island'' and trapped ``lagoon''}
\label{ssec:atoll}
\begin{figure}[t]
	\centering
	\includegraphics[width=0.5\textwidth]{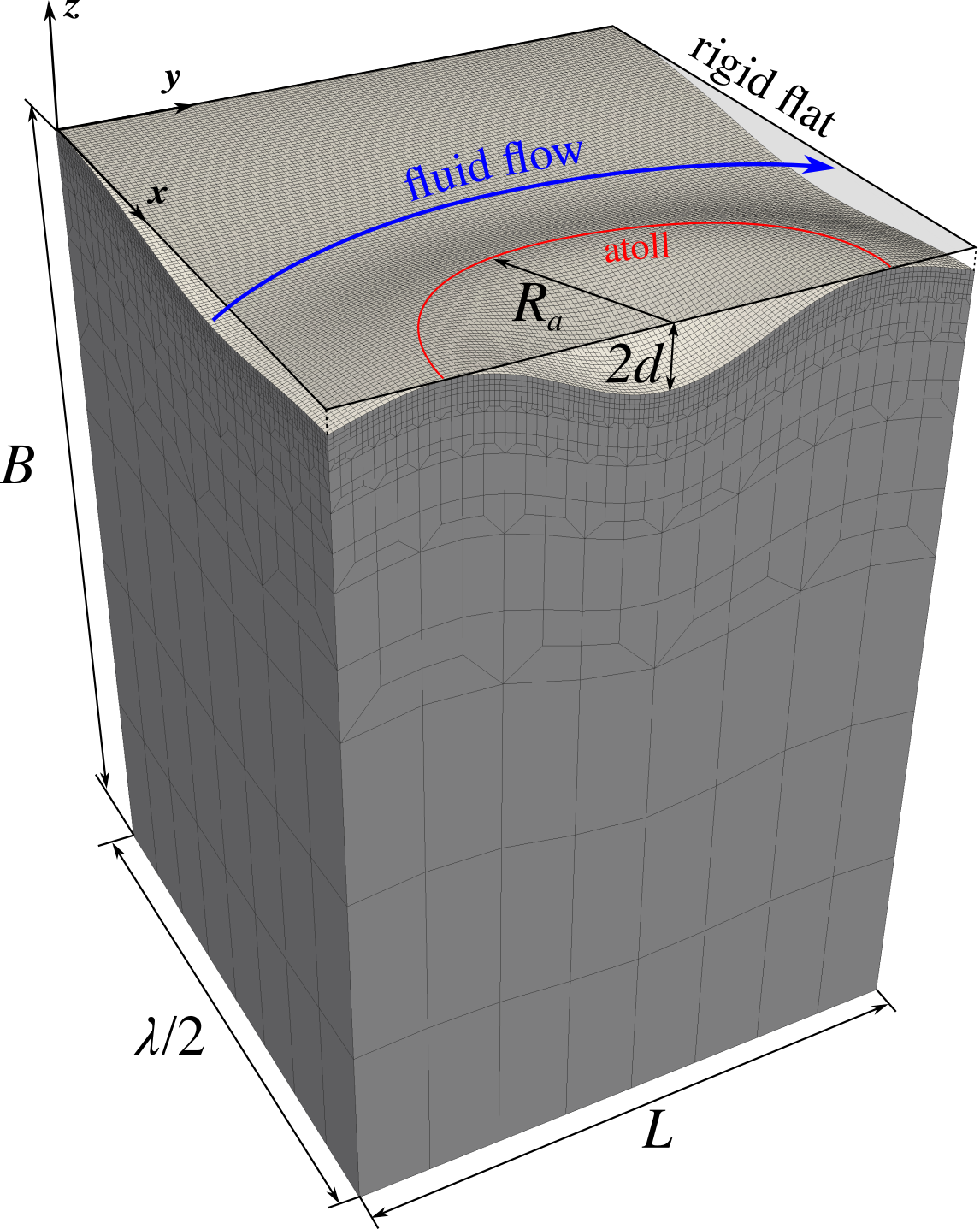} 
	\caption{Finite-element mesh with $128\times128$ faces in the contact interface which was used to solve the problem of the fluid flow across a channel with an ``atoll'' island. Note that the amplitude of the surface profile is exaggerated to highlight its features, while in the actual simulation the following geometrical parameters were used: $d = 0.02\:\text{mm}, \lambda = 2\:\text{mm}, L = 1\:\text{mm}, B = 1.4\:\text{mm}, R_a = 0.33\:\text{mm}$.}
	\label{fig:atoll_mesh}
\end{figure}
We start by presenting an example of a fluid flow in the contact interface which includes also the effect of fluid entrapment. We consider the fluid flow across an extruded wavy channel with an added ring-shaped elevation of the surface at the centre of the channel. Once this surface is gradually brought in contact with a rigid flat, the elevation forms a contact patch in a shape of an ``atoll'', which encircles a ``lagoon'' where the fluid gets ultimately trapped, see Fig.~\ref{fig:atoll_mesh}. The surface of the solid is given by the formula:
\begin{align}
\label{eq:atoll}
z(x,y) &= d \left[A(x,y)\cos{\frac{2\pi x}{\lambda}}-1\right],\\
A(x,y) &= 1-2\,\frac{(x-\lambda/2)^2 + (y-L/2)^2}{R_a^2}\, e^{1-\frac{(x-\lambda/2)^2 + (y-L/2)^2}{R_a^2}},\nonumber
\end{align}
where $d$ and $\lambda$ are the amplitude and the wavelength of the channel profile, respectively, $L$ is the length of the channel, and $R_a$ is the radius of the ``atoll''. Note that the centre of the ``lagoon'' is at $(\lambda/2, L/2)$, and the atoll's elevation is equal to the the elevation of the crest on the periphery of the simulated geometry. 
In the finite-element mesh the number of regularly spaced nodes on each side of the surface is $N=129$.

On vertical faces of the solid we apply zero normal displacement: $\left.\vec{u}_y\right|_{y=0} = \left.\vec{u}_y\right|_{y=L} = 0$ and $\left.\vec{u}_x\right|_{x=0} = \left.\vec{u}_x\right|_{x=\lambda/2} = 0$, while
the bottom face is displaced vertically towards the rigid flat within 100 load steps of equal size until the fluid channel is completely sealed. We consider throughout the whole loading process a constant fluid pressures prescribed at the inlet: $\left.p\right|_{y=0} = p_\text{in}$ and the outlet $\left.p\right|_{y=L} = p_\text{out}$, complimented by conditions of zero flux at the remaining part of the boundary of the fluid domain $\Gamma^\text{f}$: $\left.q_0\right|_{x=0}=\left.q_0\right|_{x=\lambda/2}=0$. Note that due to these boundary conditions, lines $x=0$ and $x=\lambda/2$ become lines of symmetry of the problem under study.
The inlet fluid pressure is set to $p_\text{in} = 10\:\text{MPa}$ and the outlet $p_\text{out} = 0$. 

The geometrical parameters are given in the caption of Fig.~\ref{fig:atoll_mesh}. 
For the solid we consider material parameters typical for a soft matter which can be used in sealing applications: Young's modulus $E = 1\:\text{GPa}$ and Poisson ratio $\nu = 0.4$ (effective elastic modulus is $E^* = E / (1-\nu^2)\approx 1.19\:\text{GPa}$), while fluid parameters are of a typical mineral oil with initial bulk modulus $K_0 = 2\:\text{GPa}$ and $K_1 = 9.25$~\citep{kuznetsov1985effect}. Note that due to considered zero-flux boundary conditions and the form of the Reynolds equation~\eqref{eq:reynolds}, the fluid pressure field and induced tractions do not depend on the actual value of the fluid viscosity, therefore, we used an arbitrary value of $\mu = 1.0\:\text{Pa}\cdot \text{s}$.

\begin{figure}[p]
	\centering
	\includegraphics[width=0.99\textwidth]{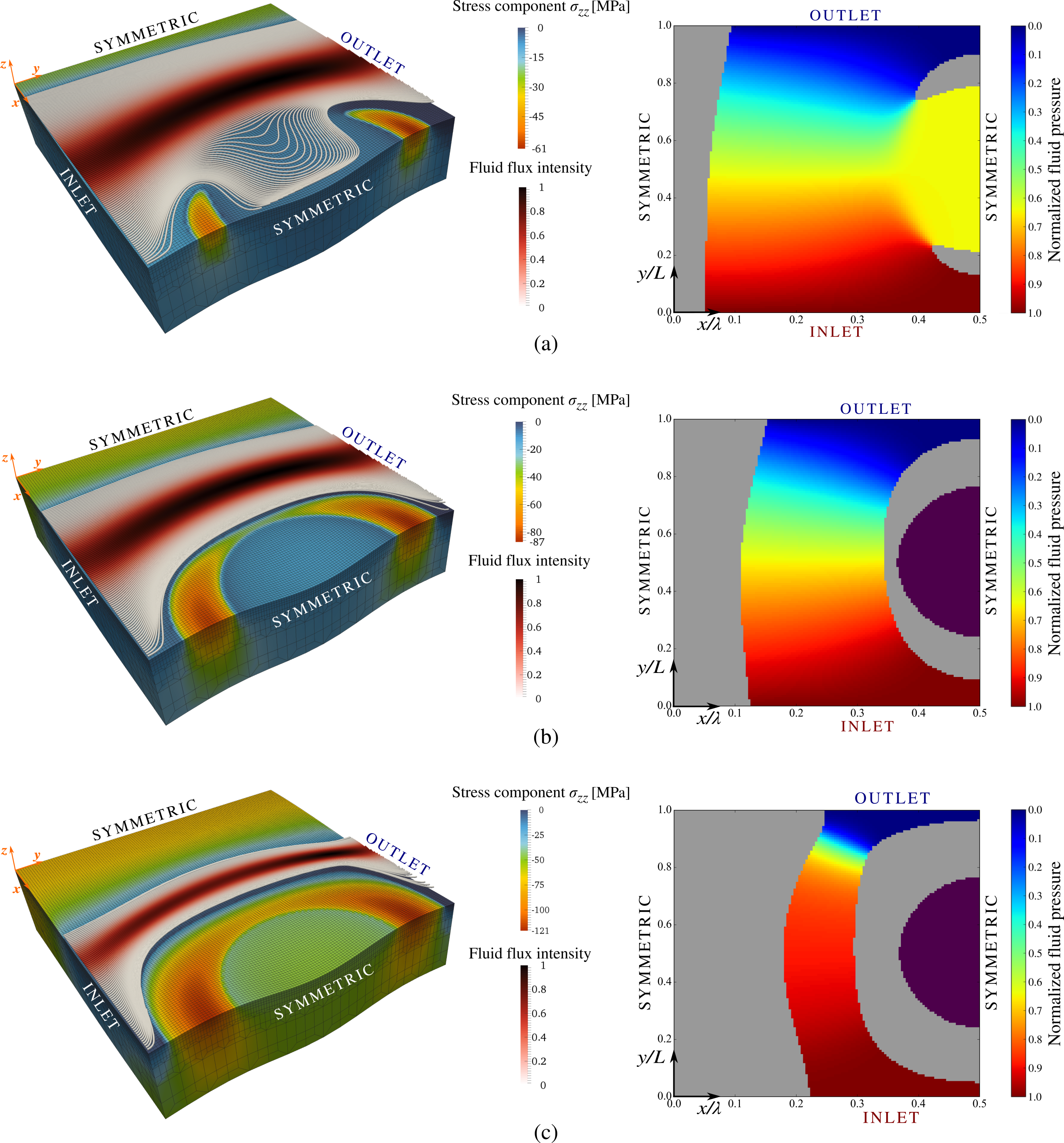} 
	\caption{Fluid flow across a wavy channel with an ``atoll'' island. Three different load steps are presented: (a) beginning of the loading, contact occurs only at the crest of the channel's profile and at two opposite sides of the atoll, therefore the fluid is not trapped yet (applied pressure is $p_\text{ext}/E^*\approx0.007$); (b) the two, initially separated, atoll's contact zones merge and fluid is trapped ($p_\text{ext}/E^*\approx0.015$); (c) under the increasing load, the trapped fluid is further pressurized ($p_\text{ext}/E^*\approx0.037$). For each loading step in the left column the bulk view of the solid is shown, with colour representing the $\sigma_{zz}$ component of the stress tensor, moreover, fluid flow lines with the colour representing the normalized fluid flux intensity $q/q_\text{max}$ are added. In the right column the interface view is given, with colour representing the normalized fluid pressure in the flow $p/p_\text{in}$, the contact patches are shown in grey colour and the trapped fluid zone is purple. Note that the trapped fluid pressure corresponding to the loading step (b) is $p^\text{tf}/p_\text{in} \approx 1.2$, step (c): $p^\text{tf}/p_\text{in} \approx 4.5$. The full animated solution can be found in Supplementary material~\citep{supp}.}
	\label{fig:atoll_bulk}
\end{figure}

Results of the simulation are summarized in Fig.~\ref{fig:atoll_bulk}, note that only three of 100 load steps are shown. 
The full animated solution can be found in Supplementary material~\citep{supp}.
At the beginning of the loading, see Fig.~\ref{fig:atoll_bulk}(a), the atoll's contact zone grows from two opposite (inlet and outlet) sides, therefore the fluid is not yet trapped and can freely flow through the forming ``lagoon''. At the same time a contact zone on the crest at $x=0$ also starts to grow from the outlet side. Note that all contact zones are not symmetric with respect to the line $y = L/2$ (which would be the case if one-way coupling was studied for the same set-up), since the fluid pressure applied to the surface of the solid is maximal at the inlet and is monotonically decreasing towards the outlet. The fluid pressure inside the lagoon before its closure is almost uniform and is higher than the mean value between the inlet and outlet fluid pressures. This is also an effect of considering the two-way coupling, since the fluid pressure distribution from the inlet to the outlet is concave (contrary to a linear distribution along the $OY$ axis in the case of one-way coupling), as was also shown in~\citep{shvarts2018fluid}. At the second stage, see Fig.~\ref{fig:atoll_bulk}(b), corresponding to higher external load, two atoll's zones of contact merge and form a non-simply connected patch, which encircles a non-contact area inside the lagoon with trapped fluid. Under increasing external pressure, see Fig.~\ref{fig:atoll_bulk}(c), the contact area continues to grow, reducing the area corresponding to the fluid flow. At the same time, the area of the trapped fluid zone is decreasing much slower due to a higher fluid pressure in the lagoon than in the fluid-flow zone, which can be observed in the figure by comparing the value of stress component $\sigma_{zz}$ at $\Gamma^\text{fsi}$ (bottom of the channel) and at $\Gamma^\text{tf}$ (bottom of the lagoon). 

\subsubsection{Comparison of the interface transmissivity between one- and two-way coupling approaches}

In order to compare one- and two-way coupling approaches and demonstrate the effect of the trapped fluid, we compute an integral value: the transmissivity of the interfaces for a given external load, which is a measure of the volume of the fluid passing through the interface in a unit time under a unit pressure drop. We use the following definition of the \textit{effective} transmissivity:
\begin{equation}
K_{\text{eff}} = -\frac{12\mu Q L}{d^3 (p_\text{out}-p_\text{in}),}
\end{equation}
where $Q$ is the mean flux over the area $\lambda/2 \times L$, i.e. 
\begin{equation}
\label{eq:mean_flux}
Q=\frac{2}{\lambda L}\int\limits_{0}^{\lambda/2}\int\limits_{0}^L q_y\, dxdy,
\end{equation}
and $q_y =  -g^3/(12\mu)\:\partial p / \partial y$ is the fluid flux in the $y$-direction.

In Fig.~\ref{fig:atoll_transmis}(a) we show that the transmissivity of the interface for a given external load is lower under the one-way coupling, which neglects the effect of fluid pressure on the surface of the solid, than in the two-way coupling. Accordingly, the critical pressure needed to seal the channel is ${\approx}17\%$ higher if the two-way coupling is considered, which is an agreement with~\citep{shvarts2018fluid}, where we found the critical sealing pressure to be an affine function of the inlet pressure for the case of a wavy channel without trapped fluid zone.  
Moreover, once a trapped fluid pool is formed, it provides additional load-bearing capacity, since its pressure is increasing with the increasing external load. 
Therefore, the observed critical sealing pressure is further increased, becoming ${\approx}20\%$ higher than the value obtained under the one-way coupling. Note also that in the same figure we plotted the curve of the transmissivity evolution
under one-way coupling shifted by an external pressure offset $p_\text{in}/2$. This
offset represents a simple estimation of the additional fluid-induced load-bearing capacity,
which is accounted for in the two-way coupling approach (we recall that $p_\text{out} = 0$).
The shifted curve coincides with the transmissivity of the two-way coupling in the region of intermediate loads and underestimates the transmissivity closer to the complete sealing, predicting only ${\approx}11\%$ higher critical load. Moreover, the simple estimation of the load-bearing capacity cannot take into account the effect of the trapped fluid, which was shown to further increase the transmissivity in the numerical simulation.

Additionally, in Fig.~\ref{fig:atoll_transmis}(b) we plot the effective transmissivity as a function of the fraction of real contact area for the same three simulation as in Fig.~\ref{fig:atoll_transmis}(a). Here, curves for one- and two-way coupling neglecting the effect of the trapped fluid almost coincide, which is also in agreement with previous results~\citep{shvarts2018fluid}. However, the curve corresponding to the case accounting for trapped fluid is different from two other cases and shows ${\approx} 8\%$ smaller contact area at the complete sealing. It is important to note, that this difference is not equal to the area of the trapped zone $A_\text{t}/A_0 \approx 0.11$ (which is out of contact), but reflects the ``resistance'' to the contact area growth caused by the pressurized trapped fluid. Note also that here the refined approach to the contact area computation (see Section~\ref{sec:contact_area}) was used.

\begin{figure}[t]
	\centering
	\includegraphics[width=1.0\textwidth]{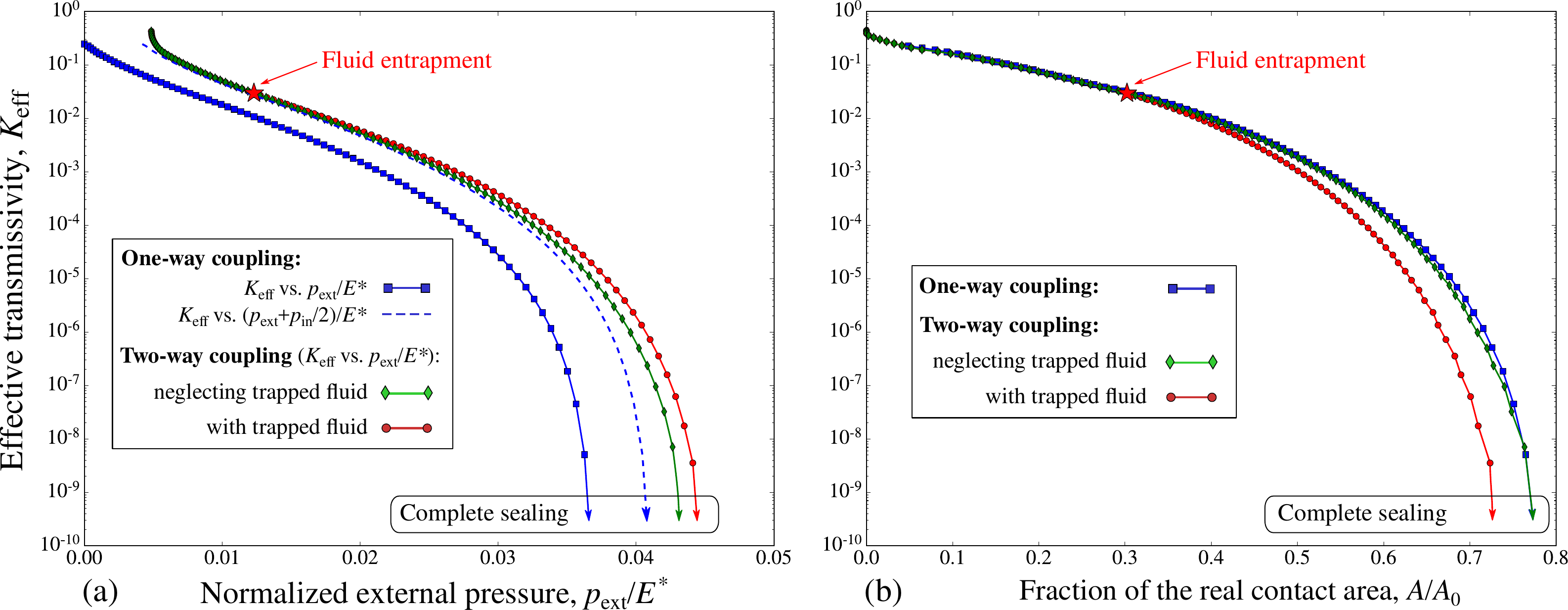} 
	\caption{(a) The evolution of the effective interface transmissivity $K_\text{eff}$ under increasing external load comparing three simulations: 
	one-way coupling approximation, two-way coupling neglecting the effect of trapped fluid and two-way coupling accounting for the trapped fluid. Additionally, the results of the one-way coupling simulation are shown shifted by an offset of the external load equal to $p_\text{in}/2$. (b) Effective transmissivity as a function of the real contact area fraction for the same three simulation as in (a). In both sub-figures the star symbols denote the event of fluid entrapment.}
	\label{fig:atoll_transmis}
\end{figure}

\subsubsection{Convergence of the Newton-Raphson method}

\begin{figure}[t]
	\centering
	\includegraphics[width=1.0\textwidth]{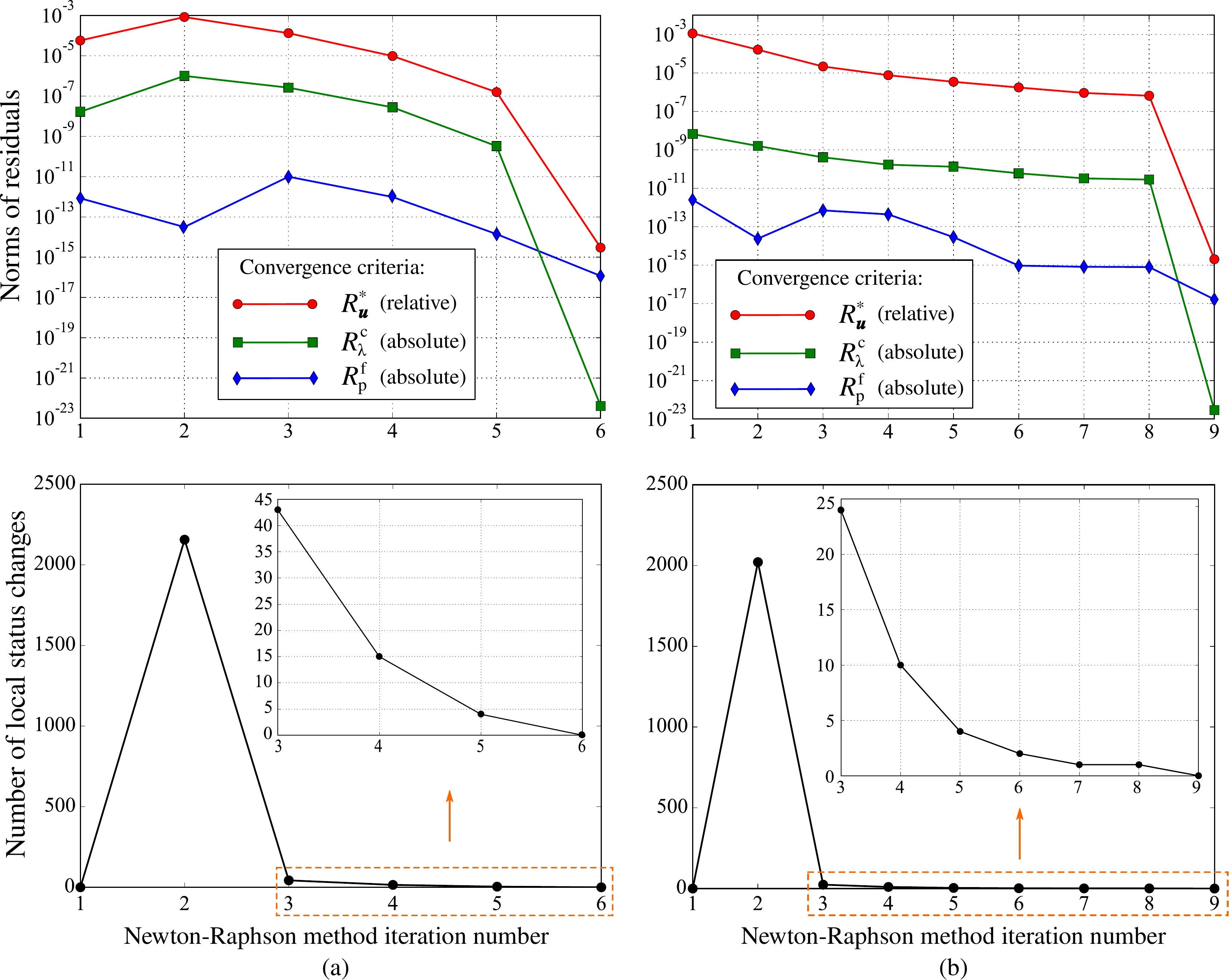} 
	\caption{Residual-wise (top) and status-wise (bottom) convergence of the Newton-Raphson method corresponding to a load step during which the trapped fluid zone is formed. Results presented for 2 simulations with different  values of the fluid inlet pressure: (a) $p_\text{in} = 2\:\text{MPa}$ and (b) $p_\text{in} = 10\:\text{MPa}$. Residual-wise convergence is shown in terms of the relative criterion for the residual corresponding to the displacement DOFs and the absolute criterion for the other two residuals, see~\eqref{eq:newton_cond}. The status-wise convergence is shown for the number of status changes defined in~\eqref{eq:status_change}. Note that the evolution of the number of status changes from the 3rd iteration until the last one is shown in insets for each considered case.}
	\label{fig:newton_raphson}
\end{figure}

We demonstrate in Fig.~\ref{fig:newton_raphson} the residual-wise and the status-wise convergence of the Newton-Raphson method corresponding to one particular load step, 
during which two atoll's contact patches merge and encircle the trapped fluid zone. 
Note that this step is the most challenging of the whole sequence from the convergence point of view, since the highest number of local status changes is observed during this step. 
We present for comparison results obtained in 2 simulations with different values of the fluid inlet pressure: $p_\text{in} = 2\:\text{MPa}$, see  Fig.~\ref{fig:newton_raphson}(a), and $p_\text{in} = 10\:\text{MPa}$, see Fig.~\ref{fig:newton_raphson}(b), while for both cases $p_\text{out} = 0$. We also used in both simulations the following tolerance thresholds for the three residuals: $\epsilon_{\vec{u}} = 10^{-6}, \epsilon_\lambda = \epsilon_p = 10^{-12}$ (MPa). 

According to presented results, while the local status of faces keeps changing between iterations, see~\eqref{eq:status_change}, the residual-wise convergence is not in general quadratic. 
Note that a high peak of the number of status changes, corresponding to the 2nd iteration in both cases, is caused by the first detection of the trapped fluid zone. 
However, the number of status changes monotonically decreases after the 2nd iteration. 
Eventually, the quadratic residual-wise convergence is achieved when the number of status changes reaches zero value in case of $p_{\text{in}} = 10\:\text{MPa}$ (and even one iteration before that for $p_{\text{in}} = 2\:\text{MPa}$), implying that the correct partition of the interface for the given external load has been found. In case of a lower fluid pressure, see Fig.~\ref{fig:newton_raphson}(a), the error drops below the specified tolerance  after 6 iterations. However, in case of higher fluid pressure, see Fig.~\ref{fig:newton_raphson}(b), more iterations are needed to find the correct status for each face. Note also that the quadratic residual-wise convergence is observed only for displacement DOFs and the Lagrange multipliers, while the norm of the fluid-flow residual $||\vec{R}^\text{f}_p||$ by this iteration is already below the convergence threshold $\epsilon_p$ and even close to the double machine precision ${\approx}10^{-16}$.

We would like to comment on the choice of the augmentation parameter $\epsilon$ required for the solution of the contact sub-problem (see Section~\ref{ssec:contact}), which in theory does not affect the solution, but may have effect on the convergence. From our experience, the choice of $\epsilon$ is problem-dependent, and for a given set-up, a numerical experiment might be required to find an optimal value. Following~\citep{cavalieri2013augmented}, we suggest $\epsilon = 0.01\,E^*/h^3$ (where $h$ is the smallest length of an edge in the mesh) as the initial choice, which can be later adjusted depending on the observed effect on the convergence. Note that in all considered examples in this study we used $\epsilon = 10^8$ N/mm$^5$. 

Moreover, our study showed an unusual dependence of results on the augmentation parameter, which was, however, rather weak. A small oscillation of the surface traction field appeared in the solution at the border between the contact and the fluid flow and/or trapped fluid zones, see Fig.~\ref{fig:atoll_bulk}(c), it can be also observed in Fig. 6-8 in~\citep{shvarts2018fluid}. We speculate that the reason for this oscillation is the piecewise-continuous interpolation of the contact Lagrange multipliers~\eqref{eq:interp}, while in theory the contact pressure field has less regularity, see Section~\ref{sec:weak:cont}. Therefore, it is expected that the use of dual shape functions~\citep{popp2013improved} for interpolation of Lagrange multipliers can remove this oscillation. At the same time our studies showed that the oscillation is decreased if the value of the augmentation parameter is increased. Unfortunately, the augmentation parameter $\epsilon$ cannot be arbitrarily high, since it may lead to poor conditioning of the global tangent matrix. However, the discussed artefact does not affect the solution in the whole domain, and therefore does not undermine the consistency of the proposed method. 
Nevertheless, it presents an interesting topic of a future investigation.

\subsection{Fluid flow through a representative rough contact interface}
\label{ssec:rough}

\begin{figure}[t]
	\centering
	\includegraphics[width=1.0\textwidth]{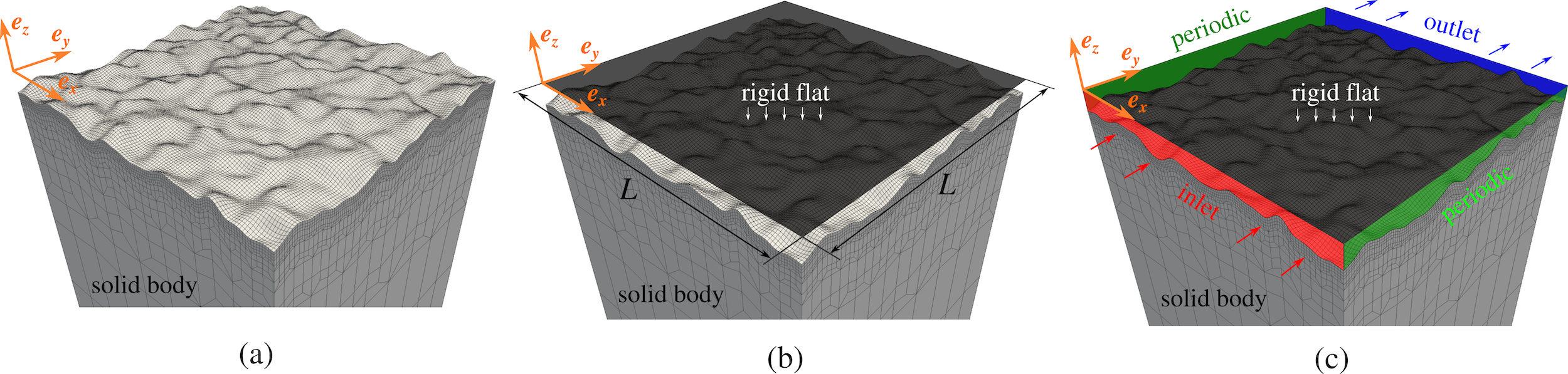} 
	\caption{Sketch of the second problem under study: contact between a deformable solid with a periodic representative rough surface (a), brought in contact with a rigid flat (b), in presence of the thin fluid flow in the free volume between the two surfaces (c). Note that the amplitude of the surface roughness is exaggerated, while in the actual simulation we used a surface with root mean squared (rms) of heights $S_q = 1\:\mu \text{m}$ and rms of the height gradient $S_{dq}\cong 0.055$. The lateral size of the studied square surface is $L=1\:\text{mm}$, while the  vertical size of the FEM mesh $B = 1.4\:\text{mm}$ is the same as in the first example.}
	\label{fig:sketch_rough}
\end{figure}

\begin{figure}[p]
	\centering
	\includegraphics[width=0.99\textwidth]{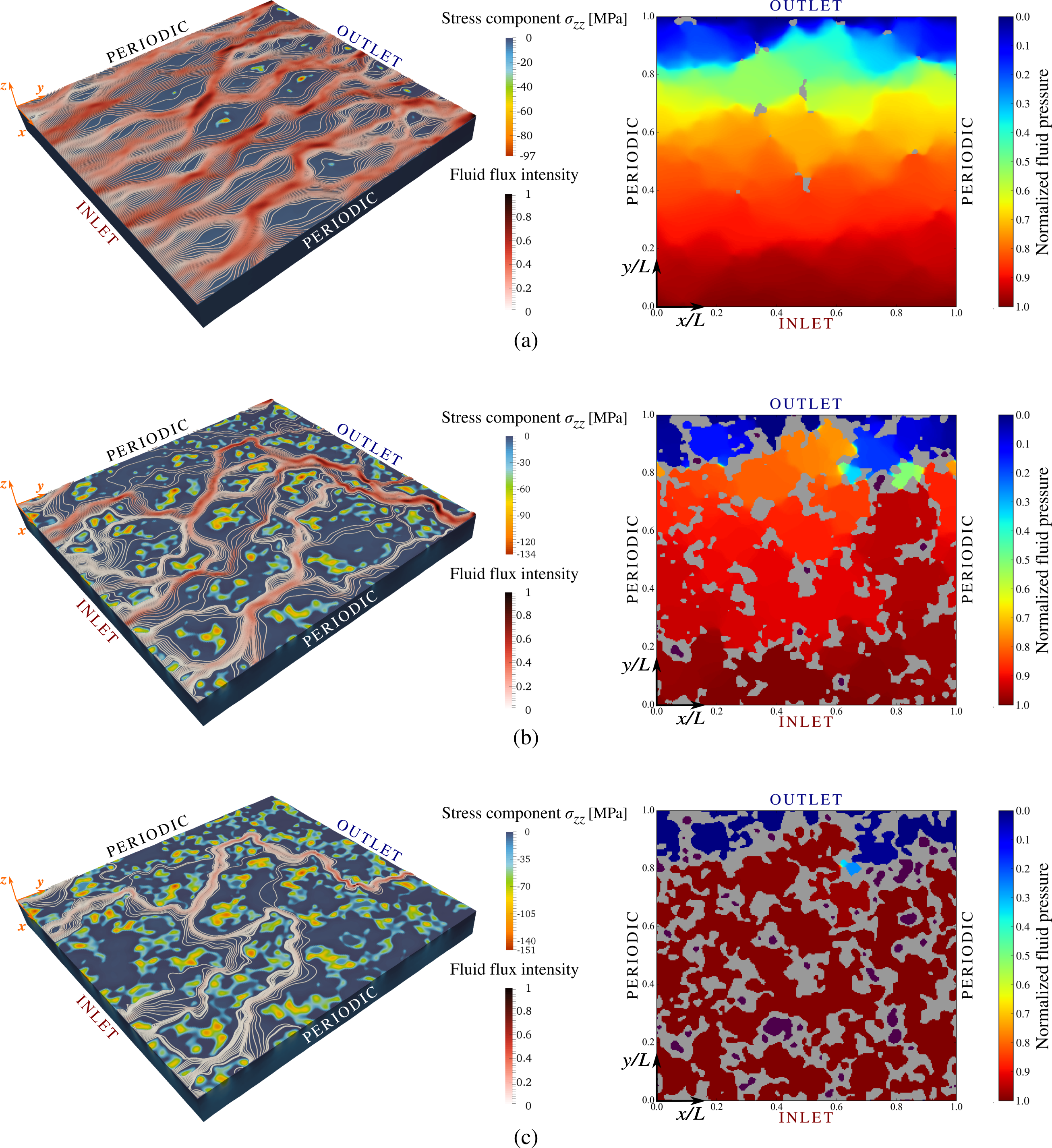} 
	\caption{Fluid flow through the contact interface between a deformable solid with representative rough surface and a rigid flat. Three different load steps with increasing external pressure are presented: (a) $p_\text{ext}/E^*\cong0.002$, (b)  $p_\text{ext}/E^*\cong0.007$, (c)  $p_\text{ext}/E^*\cong0.013$. For each loading step in the left column the bulk view of the solid is shown, with colour on the surface representing $\sigma_{zz}$ component of the stress tensor, moreover, fluid flow lines with the colour representing the normalized fluid flux intensity $q/q_\text{max}$ are added. In the right column the interface view is given, with colour representing the normalized fluid pressure in the flow $p/p_\text{in}$, the contact patches are shown in grey colour and all trapped fluid zone are purple (note that the fluid pressure in each trapped zone is different and is  increasing with the increasing external loading). At the step (b) $9$ trapped fluid zones are present ($n_\text{tf} = 9$), the highest trapped fluid pressure is $p^\text{tf}/p_\text{in} \cong 3.6$, at the step (c) $n_\text{tf} =54$, highest pressure $p^\text{tf}/p_\text{in} \cong 6.4$. The full animated solution can be found in Supplementary material~\citep{supp}.}
	\label{fig:rough_bulk}
\end{figure}

The second example is a problem of contact between a deformable solid with a representative \textit{rough} surface and a rigid flat in presence of thin fluid flow in the free volume between the two surfaces, see Fig.~\ref{fig:sketch_rough} for the sketch of the problem set-up. A physically relevant simulation of a rough surface requires a very fine discretization, which becomes a bottleneck in FEM studies. Therefore, a commonly used approach is to model a part of the surface, which is small enough to make the computation possible, and at the same time large enough to act as a representative surface element, see~\citep{yastrebov2011cr} for details.

Furthermore, the spectrum of surface roughness should be sufficiently rich to be physically representative at least to a certain extent, i.e. the frequency cut-offs in the model spectrum need to be chosen with some physical motivation and kept at values for which the continuum mechanics remains to be a valid approximation, see, for example~\citep{robbins2005n,solhjoo2019nanocontacts}. 
Using the filtering approach~\citep{hu1992ijmtm} also discussed in~\citep{yastrebov2015ijss}, we generated a periodic surface of a period $L$ with the following parameters: smallest wavenumber $k_l = 8\pi/L$ (which corresponds to the longest wavelength $\lambda_l=L/4$), highest wavenumber $k_s=64\pi/L$ (corresponding to the shortest wavelength $\lambda_s = L/32$), number of points on each side of the surface is $N=257$, Hurst exponent is $H=0.8$. 
Similarly to the first example, we consider throughout the whole loading process a constant fluid pressures prescribed at the inlet: $\left.p\right|_{y=0} = p_\text{in}$ and the outlet $\left.p\right|_{y=L} = p_\text{out}$. However, following the approach of a representative surface element, here we consider periodic boundary conditions at two other sides of the fluid domain: $\left.p\right|_{x=0}=\left.p\right|_{x=L}$. On lateral vertical faces of the solid adjacent to the inlet and the outlet zones we apply the boundary conditions of zero normal displacement $\left.\vec{u}_y\right|_{y=0} = \left.\vec{u}_y\right|_{y=L} = 0$, while on two other faces we prescribe the periodic boundary condition: $\left.\vec{u}\right|_{x=0} = \left.\vec{u}\right|_{x=L}$. The bottom face of the solid is displaced vertically towards the rigid flat within 100 equally-sized load steps until all channels connecting the inlet and the outlet are closed. We use here the same material properties as in the previous example, while the geometrical parameters are given in the caption of Fig.~\ref{fig:sketch_rough}.

Results of the simulation with the fluid inlet pressure $p_\text{in} = 4\:\text{MPa}$  and zero outlet pressure are presented in Fig.~\ref{fig:rough_bulk}, note that only 3 load steps out of 100 are shown. 
The full animated solution can be found in Supplementary material~\citep{supp}.
At the first shown step (a) trapped fluid zones are not yet observed, however, they  appear during further loading. Interestingly, atoll-type zones, which were studied previously in the model geometry, appear naturally in case of a representative rough interface, see Fig.~\ref{fig:rough_bulk}(b)~and~(c),  at the last presented step the number of these zones reaches $n_\text{tf} = 54$. 
%
It is also important to note, that the spatial distribution of the fluid pressure on its way from the inlet to the outlet changes drastically with the increasing external load. 
In Fig.~\ref{fig:rough_bulk}(a) the pressure decreases rather gradually and smoothly over the whole interface, with a more rapid change closer to the outlet, which is an effect of considering the two-way coupling. 
However, under a higher external load, see Fig.~\ref{fig:rough_bulk}(b), we observe a certain ``clusterization'' of the fluid pressure field, which becomes divided into zones partially surrounded by contact patches. 
Within these ``clusters'' the fluid pressure varies little and the intensity of the fluid flow is low, however, the fluid pressure gradient in narrow channels connecting these ``clusters'' is high. Under the external pressure close to the complete sealing of the interface, see Fig.~\ref{fig:rough_bulk}(c), the major part of the fluid-flow domain is under the inlet pressure, while almost all remaining part is under the outlet pressure, and virtually all pressure drop is happening over a narrow constriction connecting these two zones, which is in agreement with theoretical predictions~\citep{persson2008theory} and previous numerical simulations~\citep{dapp2016fluid}, performed under the one-way coupling approach.

\subsubsection{Comparison of the interface transmissivity}
\label{sec:comp_area}

Similarly to the first example, we compare the effective transmissivity of the contact interface between the representative surface element and a rigid flat in case of one-way and two-way coupling approaches. Considering the random surface roughness we compute the effective transmissivity as
\begin{equation}
K_{\text{eff}} = -\frac{12\mu Q L}{S_q^3 (p_\text{out}-p_\text{in})},
\end{equation}
where $Q=\langle q_y\rangle$ is the mean flux over the apparent contact area $A_0 = L \times L$, and $S_q = \sqrt{\langle (z-\langle z\rangle)^2\rangle}$ is the rms of surface heights, where
\begin{equation}
\label{eq:mean_flux_2}
\langle \bullet \rangle=\frac{1}{L^2}\int\limits_{0}^{L}\int\limits_{0}^L \bullet\, dxdy.
\end{equation}
In this study we consider $p_\text{in} = 5\:\text{MPa}$, while $p_\text{out}$ is zero.
\begin{figure}[t]
	\centering
	\includegraphics[width=0.99\textwidth]{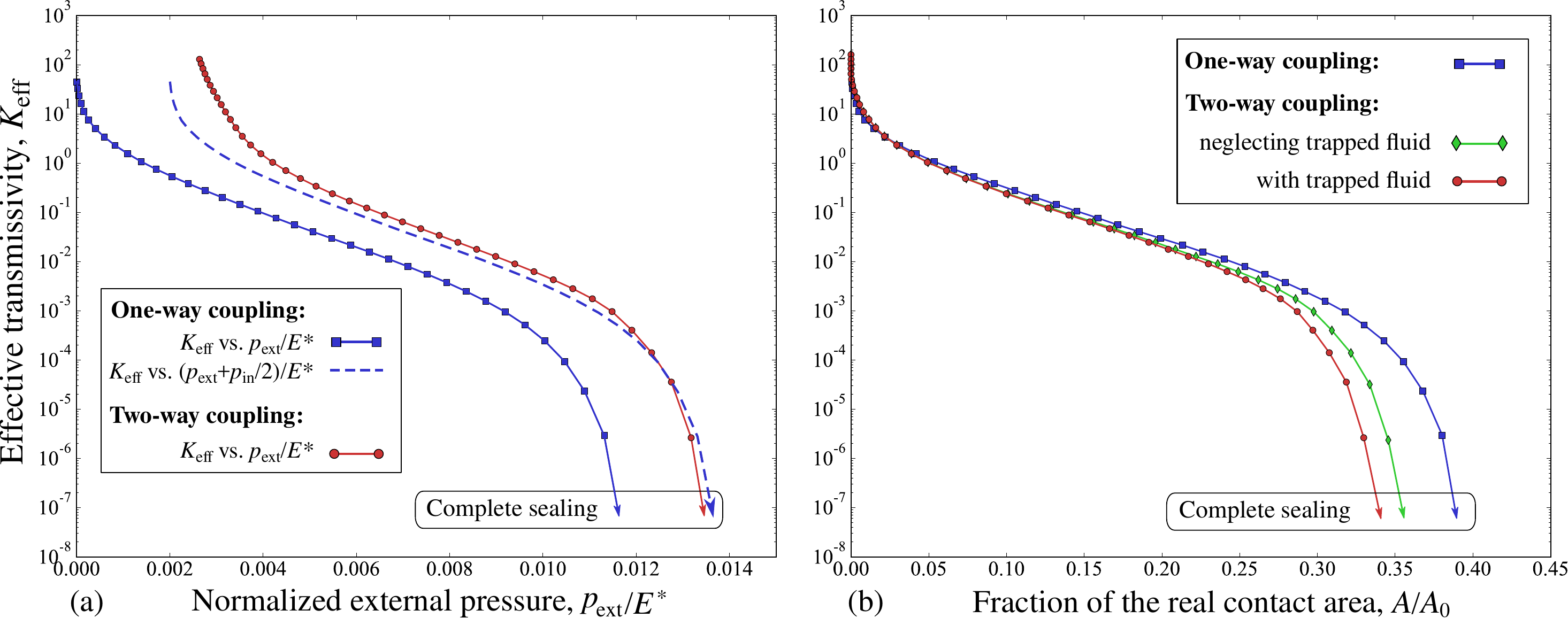} 
	\caption{(a) The evolution of the effective interface transmissivity under increasing external load comparing one- and two-way coupling approaches. Additionally, the results of the one-way coupling simulation are shown shifted by an offset of the external load equal to $p_\text{in}/2$. (b) Effective transmissivity as a function of the real contact area fraction for three coupling approaches: one-way, two-way (neglected trapped fluid) and two-way (accounting for trapped fluid).}
	\label{fig:rough_transmis}
\end{figure}

In Fig.~\ref{fig:rough_transmis}(a) we present the evolution of the transmissivity under increasing external load. As in the previous example, we observe a higher transmissivity of the interface (for the same given external pressure) in case of the two-way coupling than under the one-way approach. Accordingly, the critical external pressure necessary to completely close the interface for the fluid flow is ${\approx}17\%$ higher if the two-way coupling is considered. However, in the given example the critical load is virtually the same whether the effect of the trapped fluid is considered or not. Additionally, in Fig.~\ref{fig:rough_transmis}(a) we plot the curve corresponding to the one-way coupling shifted by an offset of the external load $p_\text{in}/2$, which represents a simple estimation of the fluid load-bearing capacity in the two-way coupling approach. This estimation is in good agreement with the results of the two-way coupling simulation, except for the beginning of the loading.
However, a simple estimation cannot predict the effect of the two-way coupling on another parameter important for sealing applications: the ratio of the real contact area $A$ to the apparent one $A_0 = L\times L$, computed at the moment when the fluid flow through the interface stops, i.e. the so-called \textit{percolation threshold}.

Previous numerical studies~\citep{dapp2012prl,dapp2016fluid} based on the one-way coupling approach show that for randomly rough self-affine surfaces $A/A_0 \approx 0.42$ at the percolation. In Fig.~\ref{fig:rough_transmis}(b) we plot the effective transmissivity with respect to the fraction of the real contact area, comparing one-way and two-way coupling (the latter is considered with and without trapped fluid). The percolation threshold observed under the one-way coupling approach is $A/A_0 \approx 0.40$, which is in agreement with the aforementioned studies. However, if the two-way coupling is considered even without trapped fluid, our results show a decrease of the percolation threshold: $A/A_0\approx 0.36$. Moreover, if the effect of numerous trapped fluid pools is taken into account, the percolation threshold is further decreased down to $A/A_0\approx 0.34$. It is important to note, that the discussed difference in the real contact area is not equal to the sum of all trapped fluid zones (which are out of contact), but represents how the pressurized trapped fluid resists the contact area growth. Note also that here we used the refined approach to the contact area computation, discussed in Section~\ref{sec:contact_area}, while the comparison of the two methods of the area computations will be presented below.

Thus, both considered examples: with model geometry and with the representative random rough surface, demonstrate the capability of the developed framework for a robust simulation of a two-way coupling between fluid and solid mechanics equations. A pronounced difference of obtained results with respect to those obtained by a one-way coupling indicates the importance of using two-way coupling for physically relevant simulations.

\subsubsection{Real contact area computation\label{sec:contactareacomp}}

The real contact area and its morphology are important not only for the study of the percolation in the sealing applications, but represent the key quantity determining the interfacial behaviour in many other physical problems, see~\citep{vakis2018modeling,bowden2001b,pei2005jmps} for examples. The evolution of the ratio of real contact area to the apparent one under increasing external load determines friction, wear, adhesion, as well as electric and heat transfer through contact interfaces. 
Therefore, it is important to ensure the accurate estimation of the contact area in numerical simulations, see also~\citep{yastrebov2017accurate}. 

\begin{figure}[t]
	\centering
	\includegraphics[width=0.8\textwidth]{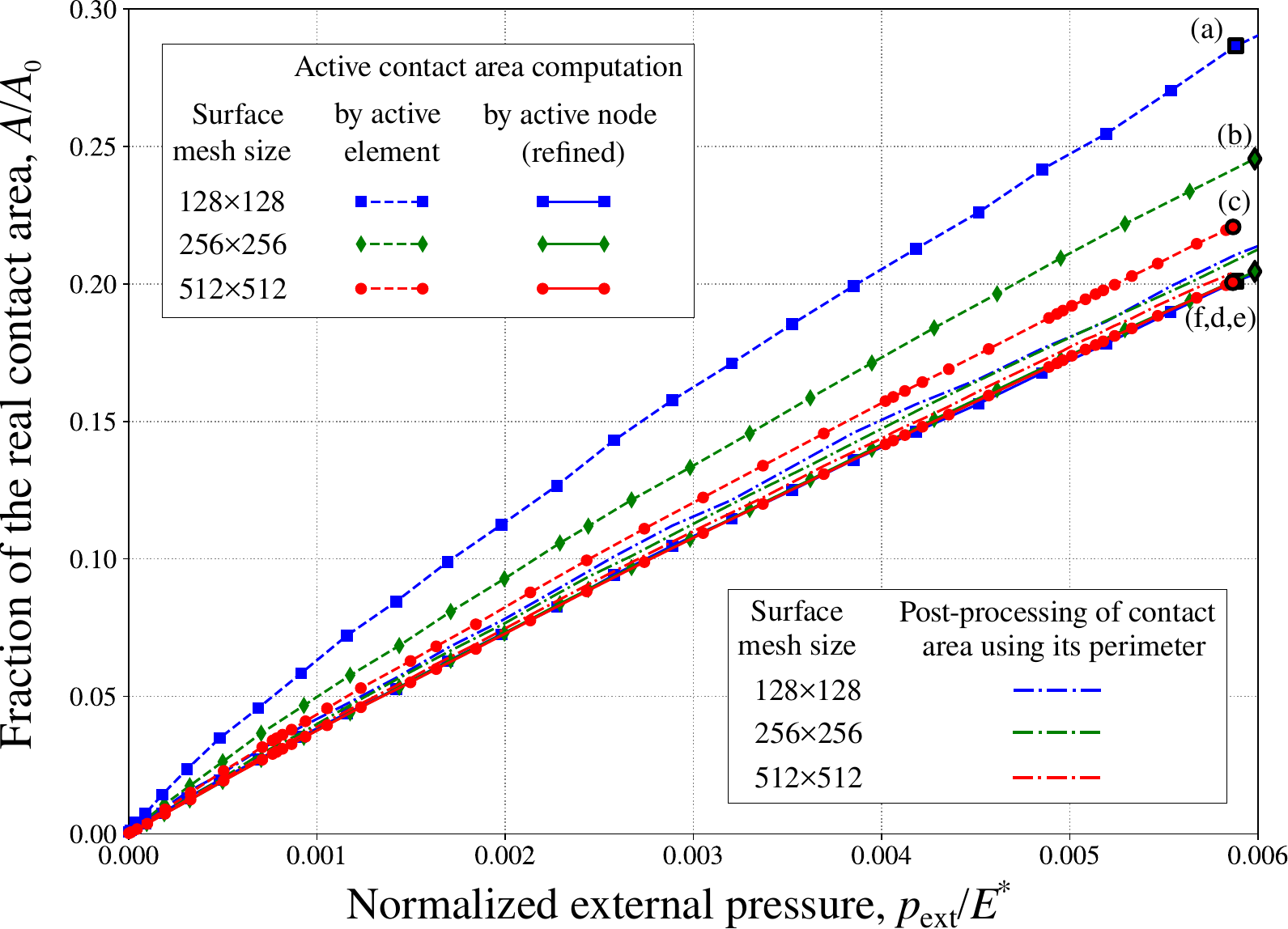} 
	\caption{Comparison of two different methods of the real area computation,~\eqref{eq:area_all} and~\eqref{eq:area_gp}. Results of simulations using three different meshes are presented (with $128\times128$, $256\times256$ and $512\times512$ face elements on the surface). For comparison, the real contact area is also calculated using correction technique based on post-processing of the contact area~\citep{yastrebov2017accurate}. The spatial distribution of the contact area for highlighted points is shown in Fig.~\ref{fig:cont_area_morph} under corresponding letters (a--f).}
	\label{fig:cont_area}
\end{figure}

We perform a mesh convergence study, comparing the values of the real contact area obtained in simulations with three different meshes: with $128\times128$, $256\times256$ and $512\times512$ face elements on the surface $\Gamma$, respectively. It is important to note that the spectrum of the surface roughness is preserved exactly the same for all considered meshes. Using the same approach as was discussed in the previous section,
we generated a surface with the following parameters: smallest wavenumber $k_l=8\pi/L$ (longest wavelength $\lambda_l=L/4$), highest wavenumber $k_s=128\pi/L$ (shortest wavelength $\lambda_l=L/64$), Hurst exponent $H=0.8$.  
The generated roughness with $513\times513$ points is mapped on the corresponding mesh. In order to obtain the surface geometry with coarser discretization ($257\times257$ and $129\times129$ points), a point-wise sampling was used, which is easy to perform, since the generated finite-element mesh has a regular quadrilateral grid on the surface.

\begin{figure}[t]
	\centering
	\includegraphics[width=1.0\textwidth]{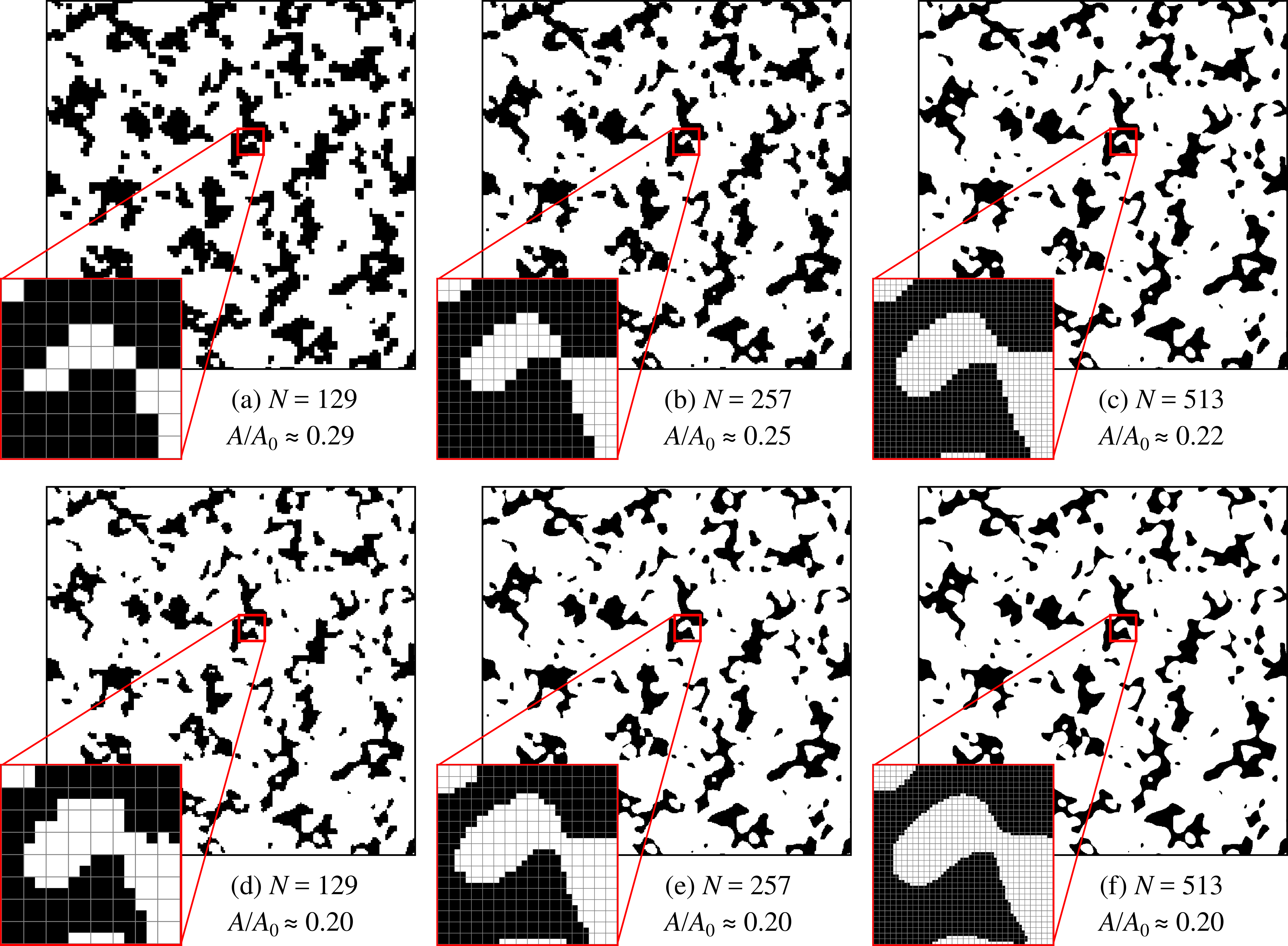} 
	\caption{Comparison of the real contact area morphology for approximately the same external load ($p_\text{ext}/E^{*}\approx0.006$) for three simulations with different mesh sizes: $N = 129$ (left column), $N=257$ (middle column), $N=513$ (right column); active contact area is shown in black, out-of-contact is white. For each case the active contact area is identified using a straightforward approach, i.e. by active element (top row), and the refined approach, i.e. by the area associated with the Gauss point closest to an active node (bottom row). A magnification of a small zone of the interface, same for all six cases, is shown in the inset of each sub-figure, with grey-coloured grid representing boundaries of contact elements.	}
	\label{fig:cont_area_morph}
\end{figure}

We present in Fig.~\ref{fig:cont_area} the comparison of the two methods of the real contact area computation, discussed in Section~\ref{sec:contact_area}. The results show a significant difference between the three meshes in case of the contact area computation based on summing up of areas of active contact elements~\eqref{eq:area_all}: the convergence seems to take place, but it is very slow. On the contrary, a refined approach to contact area computation, based on summing up for each element contribution to the contact area only from a Gauss point nearest to an active node~\eqref{eq:area_gp}, leads to a virtually mesh-independent calculation of the real contact area. For additional comparison, we plot for each of the simulations the contact area obtained using the correction technique based on post-processing of the contact area using the data on the contact perimeter, proposed in ~\citep{yastrebov2017accurate} and recently applied in~\citep{carneiro2020representative}. The results of this calculation also demonstrate convergence towards the curve corresponding to the refined approach of the contact area computation. The latter approach, however, is showing the best accuracy.

In Fig.~\ref{fig:cont_area_morph} we show the comparison of the real contact area morphology for three simulations with different mesh sizes at approximately the same external load (though not exactly the same since in these simulations we use the displacement control rather than the load control). For each mesh, both approaches of contact area identification are demonstrated: the straightforward one in (a,b,c), and the refined one in (d,e,f).
The comparison demonstrates that the refined approach provides almost the same real contact area $A/A_0\approx0.20$ for all three meshes, which is very close to the prediction of the correction technique for the finest mesh discussed above. At the same time, the results of the straightforward approach are only slowly converging towards this value of the real contact area with the increasing mesh size, and significantly overestimate it when coarser meshes are used.

However, one may erroneously expect the result of the straightforward approach for the mesh with $N=257$ and $N=513$ to be close to those of the refined approach for the mesh with $N=129$ and $N=257$, respectively (compare (b) with (d)  and (c) with (e) in Fig.~\ref{fig:cont_area_morph}) since the computation of the real contact area is performed with equal ``quanta'' of contact area, i.e. the element's area on a regular mesh is equal to the area associated with one Gauss point on a twice coarser mesh.
 Nevertheless, these computations do not coincide since even if the ``quantum'' of the contact area is the same in (b) and (d) or (c) and (e) the criterion for summing them up is different. 
 Namely, in (d) and (e) a quarter of an element is considered active if and only if the nearest node is active, while in (b) and (c), respectively, the same ``quantum'' of the contact area is added up if \textit{any} node attached to a given element is active, see also Fig.~\ref{fig:sketch_area}. Consequently, the difference in the criterion between the two approaches accounts for the overestimation of the real contact area under the straightforward approach. Therefore, even if during the numerical simulation a contact element is considered as active if at least one node is active (which is required for a consistent partitioning of the interface into contact, fluid-flow and trapped fluid zones), the refined approach for computing contact area permits to readily obtain physically more relevant values at the post-processing stage.

\section{Conclusions}

In this paper we proposed a monolithic finite-element framework aimed at solving a problem of thin fluid flow in a contact interface between a deformable solid and a rigid flat subject to a normal load. 
This framework combines a mortar-like augmented-Lagrangian-based contact resolution algorithm, 
fluid-flow elements for solving the Reynolds equation for incompressible viscous flow between immobile walls and fluid-structure interface elements to apply fluid tractions on the surface of the solid. 
Additionally, the possibility of fluid entrapment in ``pools'' delimited by contact patches and its consequent pressurization is considered using a model of compressible fluid with pressure-dependent bulk modulus. 
This model of the trapped fluid is included into the numerical framework using a superelement formulation applied separately for each trapped fluid zone. 
Furthermore, the proposed framework is suitable for both one- and two-way coupling approaches. 

One of the main difficulties of simulating the fluid flow in contact interfaces is associated with the dependency of the extent of fluid-flow domain and the trapped fluid zones on the solution of the contact problem, 
which can be enhanced by a sophisticated morphology of the contact area resulting from deterministic or random features of the surface geometry. 
In the developed framework, this complexity is handled by an on-flight procedure of partitioning the interface into contact, fluid-flow and trapped fluid zones, performed at each iteration of the Newton-Raphson method using connected-component labelling of the interface graph. The standard algorithm based on the depth-first search was further elaborated to take into account formation and evolution of trapped fluid zones.

To validate the robustness of the proposed monolithic formulation and the efficiency of the resolution algorithm, we considered a model problem with the fluid flow through an extruded wavy channel with an ``atoll''-shaped elevation of the profile, which forms one trapped fluid zone. In particular, we studied DOF-wise and status-wise convergence during the most challenging step of the whole loading sequence, corresponding to the creation of the trapped fluid zone. According to the obtained results, the number of status changes is monotonically decreasing after the 2nd iteration, and once the correct partition of the interface is found (i.e. the number of status changes between two Newton-Raphson iterations becomes zero), the quadratic convergence is achieved. Furthermore, simulations of a thin fluid flow through a contact interface between a solid with a representative rough surface and a rigid flat were demonstrated. Obtained results are in a general agreement with existing theoretical predictions and previous numerical simulations, performed using the one-way coupling approximation. However, the proposed framework permits to highlight the difference between the one-way and two-way coupling approaches, including the effect of the trapped fluid, and shows a quantitative difference in the results. Additionally, we considered a refined approach of the post-processing contact area computation, which demonstrated a practically mesh-independent result.

Using ready-to-implement formulations for all elements of the coupled problem and resolution algorithms presented in this paper, it is now possible to use the developed framework for a wide range of numerical studies featuring experimentally measured and generated surface geometries. One of possible applications of the framework is the quantification of the effect of the two-way coupling, including the phenomena of fluid entrapment, on the transmissivity of contact interfaces between solids with rough surfaces. Our preliminary results demonstrate that for  particular geometries and material parameters the accurate account for the two-way coupling can be especially important near the percolation point, which corresponds to the complete sealing of the interface.  However, the improvement of the existing understanding of coupled processes occurring in contact interfaces is important not only for sealing engineering, but is also relevant for many problems that involve thin fluid flow in narrow interfaces, appearing in biological and geophysical applications.

Finally, it is important to remark that the constructed framework could be considered as a step towards a more general methodology capable of taking into account relative tangential motion of contacting surfaces. A general computational framework for such problems at the roughness scale exists both in boundary and finite element formulations, but only in the elasto-hydrodynamic regime where the actual contact between surfaces is excluded. The case of mixed lubrication in which the boundary lubrication at contact zones competes with the elasto-hydrodynamic lubrication in non-contact zones is handled differently without direct simulation of contact zones~\citep{patir1978average}, while a unified coupled computational framework is still missing~\citep{vakis2018modeling}.

\section{Acknowledgments}

The authors are grateful to Andreas Almqvist, Francesc P\'{e}rez R\`{a}fols, Djamel Missoum-Benziane and Nikolay Osipov for enriching discussions regarding the physics of the coupled problem and the code development. The financial support of MINES ParisTech (Th\`ese-OPEN), Safran Tech, and Transvalor is greatly acknowledged.

\appendix

\section{Components of the residual vector and tangent matrix}

\label{sec:appendix}
Here we provide detailed expressions for components of residual vectors and tangent matrices formulated above for  the contact element (Section~\ref{ssec:contact}), the thin fluid flow element (Section~\ref{ssec:flow}) and the fluid-structure interface element (Section~\ref{ssec:fsi}). For all notations used here please refer to the respective Sections.

\subsection{Contact element\label{sec:app_contact}}
\begin{subequations}
	\begin{alignat}{2}
	& \mathbf{R}^\text{c}_{\vec{u}_j} = \sum\limits_{i=1}^{n}\begin{cases}
	\displaystyle \hat{\lambda}_i \, I_{ij}\frac{\partial g_j}{\partial \vec{u}_j}, & \hat{\lambda}_i\leq 0 \\
	\displaystyle 0, & \hat{\lambda}_i > 0,
	\end{cases}\\
	& \mathbf{R}^\text{c}_{\lambda_i} = \begin{cases}
	\displaystyle\tilde{g}_i, & \hat{\lambda}_i\leq 0 \\
	\displaystyle-\frac{1}{\epsilon} \lambda_i, & \hat{\lambda}_i > 0,
	\end{cases}\\
	&\mathbf{K}^\text{c}_{\vec{u}_k\vec{u}_j} = \sum\limits_{i=1}^{n}\begin{cases}
	\displaystyle \epsilon \, I_{ij} \frac{\partial g_j}{\partial \vec{u}_j} I_{ik} \frac{\partial g_k}{\partial \vec{u}_k}, & \hat{\lambda}_i\leq 0 \\
	\displaystyle 0, & \hat{\lambda}_i > 0,
	\end{cases}\\
	&\mathbf{K}^\text{c}_{\lambda_i\vec{u}_j} = \mathbf{K}^\text{c}_{\vec{u}_j\lambda_i} = \begin{cases}
	\displaystyle I_{ij} \frac{\partial g_j}{\partial \vec{u}_j}, & \hat{\lambda}_i\leq 0 \\
	\displaystyle 0, & \hat{\lambda}_i > 0,
	\end{cases}\\
	&\mathbf{K}^\text{c}_{\lambda_i\lambda_i} = \begin{cases}
	\displaystyle 0, & \hat{\lambda}_i\leq 0 \\
	\displaystyle -\frac{1}{\epsilon}, & \hat{\lambda}_i > 0,
	\end{cases} \quad \mathbf{K}^\text{c}_{\lambda_i\lambda_j} = 0 \; \text{if} \; i \ne j.
	\end{alignat}
\end{subequations}

\subsection{Fluid-flow element\label{sec:app_fluid_flow}}
\begin{subequations}
	\begin{alignat}{2}
	&\mathbf{R}^\text{f}_{p_i} = \int\limits_{-1}^1 \int\limits_{-1}^1 \left(\sum\limits_{k=1}^n N_k \: g_k\right)^3 \left( \mathbf J^{-1} \sum\limits_{j=1}^n \nabla N_j p_j \right) \left(\mathbf J^{-1} \nabla N_i\right) \det(\mathbf{J}) \, d\xi d\eta,\\
	&\displaystyle\mathbf{K}^\text{f}_{p_i p_j} = \int\limits_{-1}^1 \int\limits_{-1}^1 \left(\sum\limits_{k=1}^n N_k \: g_k\right)^3 \left(\mathbf{J}^{-1} \nabla N_i\right) \left(\mathbf{J}^{-1} \nabla N_j \right) \det(\mathbf{J}) \, d\xi d\eta,\\
	&\displaystyle\mathbf{K}^\text{f}_{\vec{u}_l p_i} = \frac{\partial g_{l}}{\partial \vec{u}_l} \int\limits_{-1}^1\int\limits_{-1}^1 3 \left(\sum\limits_{k=1}^n N_k \: g_k\right)^2 N_l \left(\mathbf{J}^{-1} \sum\limits_{j=1}^n \nabla N_j p_{j}\right) \left(\mathbf{J^{-1}} \nabla N_i \right) \det(\mathbf{J}) \, d\xi d\eta.
	\end{alignat}
\end{subequations}

\subsection{Fluid-structure interface element\label{sec:app_fsi}}

\begin{subequations}
	\begin{alignat}{2}
	&\displaystyle\mathbf{R}^\text{fsi}_{\vec{u}_i} = \sum\limits_{j=1}^n p_j \int\limits_{-1}^1 \int\limits_{-1}^1 \vec{n} \, N_i N_j \, J \, d\xi d\eta + \sum\limits_{k=1}^n \frac{g_k}{2} \int\limits_{-1}^1 \int\limits_{-1}^1 \left(\mathbf{J}^{-1} \sum\limits_{l=1}^n \nabla N_l p_{l}\right) N_i N_k \, J \, d\xi d\eta,\\
	&\displaystyle\mathbf{K}^\text{fsi}_{\vec{u}_k\vec{u}_i} = \frac{1}{2}\frac{\partial g_k}{\partial \vec{u}_k} \int\limits_{-1}^1 \int\limits_{-1}^1 \left(\mathbf{J}^{-1} \sum\limits_{l=1}^n \nabla N_l p_{l}\right) N_i N_k \, J \, d\xi d\eta,\\
	&\displaystyle\mathbf{K}^\text{fsi}_{p_l\vec{u}_i} = \sum\limits_{k=1}^n \frac{g_k}{2} \int\limits_{-1}^1 \int\limits_{-1}^1 \left(\mathbf{J}^{-1} \nabla N_l \right) N_i N_k \, J \, d\xi d\eta.
	\end{alignat}
\end{subequations}

\normalem

\end{document}